\documentclass{aa}  

\usepackage{hyperref}
\usepackage{booktabs}
\usepackage{multirow}
\usepackage{makecell}
\hypersetup{
	colorlinks=true,
	linkcolor=blue,
	urlcolor=blue,
	citecolor=blue,
}
\usepackage{graphicx}
\usepackage{txfonts}
\usepackage{lineno}

\begin{document}

	\title{Gamma-ray emission toward supergiant\\ shells in the Large Magellanic Cloud}
	
	%\subtitle{I. Overviewing the $\kappa$-mechanism}
	
	\author{Qixin~Yu\thanks{Corresponding author: qixin.yu@uibk.ac.at},
		Olaf~Reimer,
		\and
		Anita~Reimer
	}
	
	\institute{Universität Innsbruck, Institut für Astro- und Teilchenphysik, Technikerstraße 25, 6020 Innsbruck, Austria
	}
	
	\date{Received --; accepted --}
	
	% \abstract{}{}{}{}{} 
	% 5 {} token are mandatory
	\linenumbers 
	\abstract
	% context heading (optional)
	% {} leave it empty if necessary  
	{The Large Magellanic Cloud (LMC), a satellite galaxy of our Milky Way, is considered to be a very good target for investigations of diffuse gamma-ray emission. Observations have shown that diffuse emissions are related to the large- and small-scale structures of the LMC. Particularly, it has been suggested that some of the small-scale bright diffuse emissions coincide with supergiant shells (SGSs) previously identified based on H$\alpha$ images. SGSs have been proposed to be sites of powerful particle acceleration.}
	% aims heading (mandatory)
	{A first attempt is made to understand the correlation between the small-scale diffuse gamma-ray emissions and the H$\alpha$-selected SGSs. We also tried to connect them to the underlying cosmic-ray properties in these SGS regions.}
	% methods heading (mandatory)
	{We analyzed the gamma-ray emission from the LMC using more than 15 years of observations with the \textit{Fermi} Large Area Telescope in the energy range of 300$\,$MeV-300$\,$GeV. A dedicated spatial and spectral analysis was made for diffuse emission components of the LMC. Particularly, we evaluated the emissions from SGSs in detail using phenomenological approaches, and we established their corresponding spatial templates for gamma-ray emission.}
	% results heading (mandatory)
	{In the analysis of small-scale extended emissions, we detected seven H$\alpha$-selected SGSs at a significance of more than $5\sigma$. The emission structure nested inside the SGSs appears to be a non-Gaussian feature and is very likely due to a modulated cosmic-ray distribution by an ensemble of supernova remnant shocks. The spectra of the gamma-ray emission for most SGS regions favor a hadronic origin. For the large-scale diffuse emission from the LMC disk, our results suggest an interplay of hadron- and lepton-induced emission. The cosmic-ray proton energy density derived for bright SGSs is 3-10 times higher than that in the LMC disk. SGS$\,2$ was detected to be the most significant emission region among all SGSs, and only two H$\alpha$-selected SGSs were not significantly detected in this study.}
	% conclusions heading (optional), leave it empty if necessary 
	{Based on the gamma-ray excess emission we detected in H$\alpha$-selected SGSs in the LMC, we estimate that their contribution to the total diffuse gamma-ray emission in the LMC is at a non-negligible 41\% level.}
	\maketitle
	\nolinenumbers
	%
	%-------------------------------------------------------------------

\section{Introduction}\label{intro}
The evacuation of the cool interstellar medium (ISM) by combined kinetic and radiation pressure from massive stars and supernovae in HII and star-forming regions over a period of time can lead to the formation of a cavity bounded by a shell of swept-up gas in general \citep{1979ARA&A..17..213M}. The largest shells, sometimes referred to as supergiant shells (SGSs), require around $10^{53}\,\rm ergs$ or above for their creation \citep{1980MNRAS.192..365M}. It has been suggested that these SGSs with diameters of hundreds of parsec to 1~kpc driven by fast stellar winds and supernova explosions would continuously pressure the ISM onto the rims of these SGSs and thus lead to its subsequent gravitational collapse. From there, episode by episode of active star formation begins, along with the expansion of SGSs \citep{2008ApJS..175..165B}. Most energetic objects capable of accelerating cosmic rays, such as conglomerations of supernova remnants (SNRs) and superbubbles (SBs) \citep{2001SSRv...99..317B}, are expected to reside in regions with high star formation rates (SFRs)  \citep{1988AJ.....96.1874C,2016A&A...585A.162M}. SGS, as a major structure component of irregular galaxies, for which a high SFR is expected \citep{2009AJ....138.1243H}, naturally is a candidate great energetic accelerator for cosmic rays (CRs). However, the specific role played by SGSs in particle acceleration mechanisms remains unclear. Therefore, studying the physical structure of SGSs and their impact on the CR properties is crucial for us to understand the observations of them, for example, in X-ray and gamma-ray bands.  

The Large Magellanic Cloud (LMC) is an excellent target for exploring the nature of these SGSs with its well-constrained distance at $50\,\rm kpc$ \citep{2013Natur.495...76P} and low inclination angle \citep{1972VA.....14..163D}. Initially, four SGSs were identified by \cite{1978A&A....68..189G} based on an examination of H$\alpha$ images of ionized gas in HII regions. The number of identified SGSs later rose to nine \citep{1980MNRAS.192..365M}. The LMC is considered one of the brightest sources of diffuse gamma-rays detected outside the Milky Way \citep{1998ApJ...494..523S}. As CRs propagate through the LMC, they lose energy either through diffusive escape from the galaxy if sufficiently energetic, or through radiative processes and particle–particle collisions. Diffuse gamma-ray emission is then often produced in this way via leptonic and/or hadronic mechanisms. Particularly, it has been indicated that some of the small-scale diffuse gamma-ray emission regions might correlate with SGSs, and simultaneously, with massive star-forming regions traced by ionized gas in the LMC \citep{2010A&A...512A...7A}, although contributions from unresolved sources are generally expected (see \cite{2025arXiv250909028L} and the references therein). Further evidence was provided by a 73-month \textit{Fermi} Large Area Telescope (\textit{Fermi}-LAT) observation reported in \cite{2016A&A...586A..71A}, which revealed several extended regions of gamma-ray emission near or within the H$\alpha$-selected SGSs in the LMC. Their spatial distributions were modeled with Gaussian templates and classified as extended sources in the 4FGL catalog. For example, one of these extended emission regions, named E2, was found to be located within the SGS\,4, while two of the other regions (components E1 and E3) partially lie in SGS\,3 (see Fig.~\ref{fig:positions} in Appendix~\ref{app:positions}).
Although no direct TeV counterpart is reported, the nearby SB 30 Dor C \citep{2015Sci...347..406H,2024ApJ...970L..21A} was detected by H.E.S.S. with a photon index of $\sim2.6$ and an energy flux of $\sim4\times10^{-13}\,\rm erg\,cm^{-2}\,s^{-1}$ at 1\,TeV. Its emission may be embedded in components E1 and E3 as 30 Dor C is not detected by \textit{Fermi}-LAT. Furthermore, the spatial extension of 30 Dor C ($\sim0.038^\circ$) is smaller by about ten times than the size of components E1 and E3. Similarly, the other two TeV emitters observed by H.E.S.S., namely N157B (associated with 4FGL\,J0537.8-6909) and the newly detected young massive star cluster R136 (HESS\,J0538–691) \citep{2024ApJ...970L..21A}, reside in SGS\,2 and their spatial extensions are too small to significantly contribute to the emission of E1 and E3 as well. Thus, its emission nature might be attributed to other extended gamma-ray sources in the nearby region, such as SGS\,2/3.

Moreover, by studying these small-scale gamma-ray emission components from SGS regions, we were able to separate local extended emission regions from the large-scale emission of the LMC. In this way, the LMC provides us a great opportunity to compare the level of CR-induced gamma-ray fluxes and their parent CR properties in different regions as a complement to that of the Milky Way. 

The observed gamma-ray emission throughout the whole LMC is shown to be consistent with gamma-rays originating from CR interactions with the ISM and interstellar radiation fields (ISRFs) through hadronic particle-particle collision, bremsstrahlung, and inverse-Compton (IC) processes \citep{2015ApJ...808...44F}. In particular, the emission was found to have a good spatial correlation with star-forming regions traced by H$\alpha$, Wolf-Rayet stars, and SGSs, but to be poorly spatially correlated with the distribution of interstellar gas \citep{2010A&A...512A...7A}. This is seen as strong evidence supporting the conjecture that CRs are efficiently accelerated in massive star-forming regions where a significant amount of energy from strong stellar winds and supernova explosions is passed to CRs. In addition, the population of CRs in some of the small-scale gamma-ray emission regions, which perhaps correlate with SGS positions at the same time \citep{2016A&A...586A..71A}, is thought to be relatively young and more energetic than that of the large-scale regions in the LMC, implying efficient CR acceleration in these small-scale regions near or within the SGSs.

In other wavelengths, X-ray observation of the hot gas in the LMC showed that large-scale diffuse emission accounts for about 36\% of the total X-ray emission, in which $\sim6\%$ is associated with SGSs. In contrast, individual objects such as X-ray binaries and SNRs contribute over 60\% combined to the total emission \citep{2001ApJS..136...99P}. Based on this, diffuse X-ray emission of SGSs might increase significantly when they already harbor SNRs or X-ray binaries that might affect the underlying emission process. A decent fraction of SNRs is indeed detected near the rims of SGSs according to X-ray observations in the past years \citep{2016A&A...585A.162M,2024A&A...692A.237Z}. It is compelling to explore whether diffuse gamma-ray emission associated with SGSs might also constitute a non-negligible fraction of the total diffuse emission throughout the LMC.

We present a dedicated analysis of the LMC at GeV energies toward H$\alpha$-selected SGSs using more than 15 years of data recorded by \textit{Fermi}-LAT. Throughout the entire paper, the distance to the LMC is assumed to be $50\,\rm kpc$. Its structure 
is treated as a thick disk with a height of $\sim400\,\rm pc$ and a radius of $\sim3.5\,\rm kpc$ \citep{2016A&A...586A..71A}. We perform a standard \textit{Fermi}-LAT analysis in Section~\ref{sec:fermidata} and then establish alternative models to account for the high-energy observational data above 10\,GeV in Section~\ref{modeling}. We discuss the implications of our results and estimate the energy budget based on the observed gamma-ray emission levels in Section~\ref{sec:full_band}. A spectral analysis of different components is shown in Section~\ref{analysis}. Finally, we summarize and conclude in Section~\ref{sumandcon}.
%--------------------------------------------------------------------
\section{\textit{Fermi}-LAT observations and data reduction}\label{sec:fermidata}
\begin{figure*}[ht]
   \centering
   \includegraphics[width=0.33\hsize]{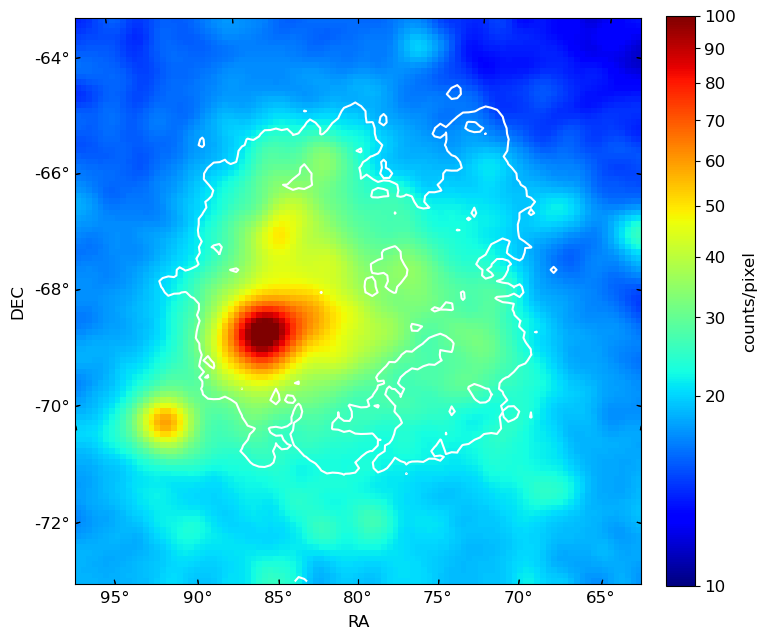}
   \includegraphics[width=0.33\hsize]{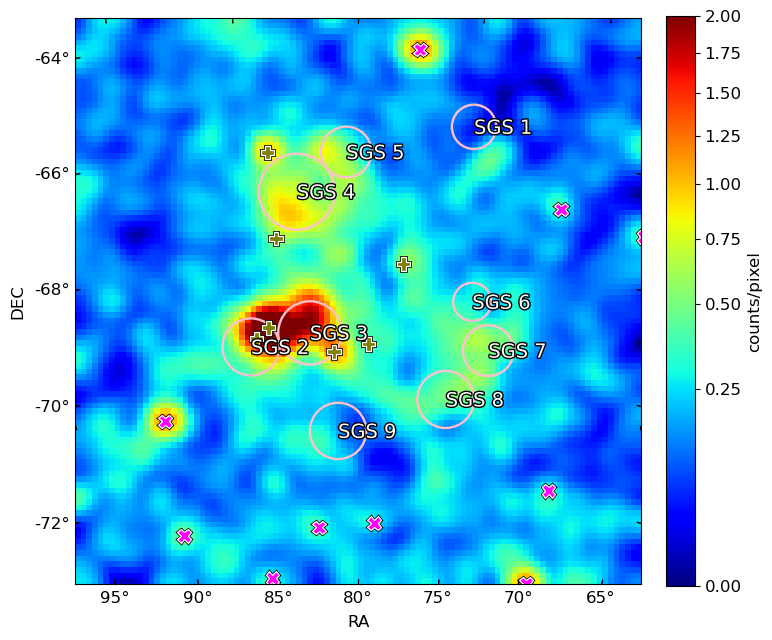}
   \includegraphics[width=0.33\hsize]{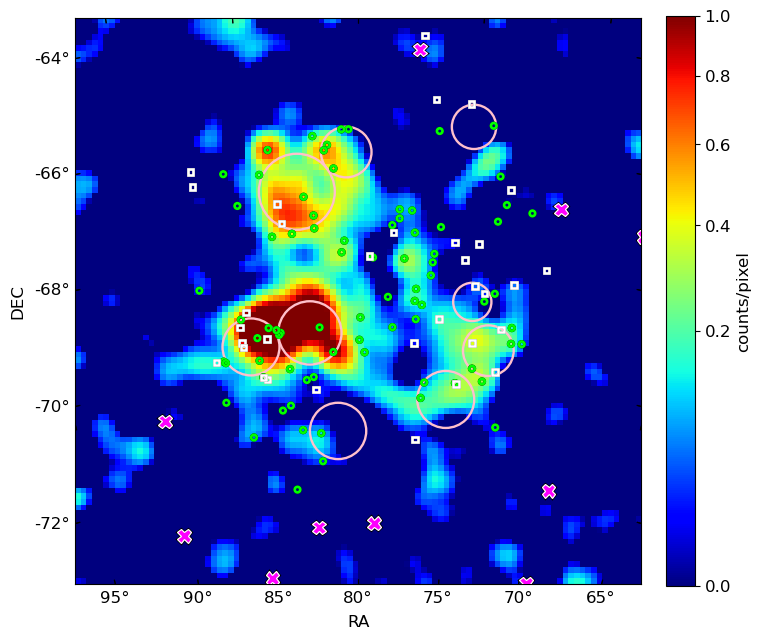}
      \caption{\textit{Left}: Observation count map of $10^\circ\times10^\circ$ ROI in the energy range of $300\,\rm MeV$-$300\,\rm GeV$. It is overlaid with the H$\alpha$ distribution contour line, which indicates a level of about 1/1000 of the peak emission in the LMC (see \cite{2003ApJS..146..407F} for H$\alpha$ data). \textit{Middle}: Observation count map of $10\,\rm GeV$-$300\,\rm GeV$. The pink circles mark the positions of each H$\alpha$-selected SGS, along with their corresponding names. Seven 4FGL-DR3 sources within the LMC boundary are plotted as olive crosses for comparisons. \textit{Right}: Residual of $10\,\rm GeV$-$300\,\rm GeV$ count map after subtracting the background catalog sources (marked with magenta crosses) and diffuse backgrounds. The residuals range from -0.23 to 3.62 in the right panel. We also plot the positions of SNR (empty green circles) and SNR candidates (empty white squares) (see \cite{2016A&A...585A.162M,2024A&A...692A.237Z} for SNR catalogs in the LMC). All maps are smoothed with a Gaussian kernel of $\sigma=0.2^\circ$ and displayed with a square-root scale for the color bar.}\label{fig:cmap}
\end{figure*}

We used more than 188 months of LAT data \citep{2009ApJ...697.1071A} taken during the period from Aug.~4 2008 to Apr.~26 2024 with an energy range of 300~MeV-300~GeV. This energy range ensures decent photon statistics and at the same time the best of LAT's angular resolution. The events were selected within a region of interest (ROI) defined as a cone centered on our target ($(\alpha_{\rm J2000},\delta_{\rm J2000})=(80.00^\circ,-68.75^\circ)$) with a half-opening angle of $10^\circ$. The data was further selected to minimize gamma-ray contribution from Earth's limb by applying a zenith angle cut of $100^\circ$. We present a smoothed count map in the left panel of Fig.~\ref{fig:cmap}, where the ROI spatial bin size is set to be $0.1^\circ\times0.1^\circ$, and the white contour outlines the H$\alpha$ distribution at a relative value of 1/1000 of the peak emission per hydrogen atoms in the LMC. For spectral analysis, we used 18 energy bins (corresponds to 6 bins per decade) throughout the paper.

The analysis was performed using \textit{Fermi Science Tools ver2.2.0}\footnote{See \url{https://github.com/fermi-lat/Fermitools-conda/wiki} for more details.} on P8R3 reprocessed data \citep{2018arXiv181011394B}. Events of class \textit{source} with conversion in front and back parts of the tracker were selected, together with the matching instrumental response functions ($\texttt{P8R3\_SOURCE\_V3}$ IRFs). The preparatory steps for data analysis including the creation of counts cubes, exposure cubes, etc., have been carried out conveniently by the Python-based package \texttt{fermipy ver1.2.2}\footnote{See \url{https://fermipy.readthedocs.io/en/v1.2/} for details.} \citep{2017ICRC...35..824W}.  

The large-scale diffuse backgrounds were modeled with templates provided by \textit{Fermi}-LAT: $\texttt{gll\_iem\_v07.fits}$ for the Galactic diffuse emission, and $\texttt{iso\_P8R3\_SOURCE\_V3\_v1.txt}$ for the extragalactic isotropic radiation. A total of 33 isolated sources are reported and listed in the \textit{Fermi}-LAT 4FGL-DR3 catalog \citep{2022ApJS..260...53A} for the region of ROI (21 sources) and $5^\circ$ outside the ROI (12 sources, to account for spillover emission inside the analyzed ROI). Of the 33 4FGL-DR3 sources, we included 22 in our background source model and excluded 11 sources located within the LMC boundary, as they might correspond to emission components that we modeled. As a result, our background source model contains 10 point sources (denoted with magenta crosses in the right panel of Fig.~\ref{fig:cmap}) that are located within the region of ROI, and the other 12 sources lie in the extended-ROI region.   

Our background model within ROI now comprises the two general diffuse background templates as mentioned above and 10 4FGL-DR3 sources in the source model. It has a total of 25 degrees of freedom, 23 of which are normalization and spectral index parameters for the 10 point sources located within the ROI while the remaining 2 correspond to the normalization parameters of the Galactic diffuse emission and extra-galactic isotropic radiation templates provided by LAT \citep{2022ApJS..260...53A}.  

\section{High-energy band: 10~GeV-300~GeV}\label{modeling}
\subsection{\textit{Fermi}-LAT spatial analysis}
For the diffuse gamma-ray analysis of the LMC, the large-scale diffuse emission across the LMC dominates in the lower energy range \citep{2010A&A...512A...7A,2016A&A...586A..71A}. Therefore, by selecting events exceeding $10\,\rm GeV$ for the spatial analysis, we can substantially suppress the contribution from diffuse gamma-ray emission due to large-scale interactions of CRs with the ISM and/or the ISRF, so that small-scale extended emissions produced by more energetic CRs in SGS regions can be analyzed. Moreover, for energies $\gtrsim10\,\rm GeV$ the LAT PSF is $\sim0.2^\circ$ (68\% containment), allowing for well-localized source locations and reduction of systematic uncertainties owing to the angular resolution in this study \citep{2009PhRvL.103y1101A}. 

The LAT analysis above $10\,\rm GeV$ follows closely the setup and steps described in Section~\ref{sec:fermidata}. A corresponding count map is presented in the middle panel of Fig.~\ref{fig:cmap}, where we also mark out the locations of H$\alpha$-selected SGSs to visually inspect if a coincidence exists between local extended emission and SGSs. This relation can be better seen in the right panel of Fig.~\ref{fig:cmap} after we subtract the background sources. Indeed, strong residuals are shown toward the SGS regions, particularly in SGSs~2-5 (see \cite{1980MNRAS.192..365M} for respective SGS parameters). 
These residual diffuse emissions are likely to be generated by interactions of accelerated CRs with the ISM and/or the ISRF inside the SGSs. This is very similar to what has been proposed in the case of SB origin of CRs \citep{1998ApJ...509L..33H,1999A&A...349..673P}. Stellar winds and supernovae in SBs combine to inject mechanical energy into the surrounding medium by means of shock waves and turbulence. This creates a viable scenario of particle acceleration supported by evidence of the SB origin of $^{22}\rm Ne$ excess in CRs, for example \citep{2003ApJ...590..822H}, numerical simulations \citep{2020MNRAS.493.3159G}, and gamma-ray observations toward the Milky Way and LMC \citep{2011Sci...334.1103A,2015Sci...347..406H,2020ApJ...893..144L}. In a sense, SGSs can be regarded as a natural extension of SBs with respect to the origin and acceleration mechanisms of CRs, although in a more dynamic and complex ISM environment. We thus present a model to describe CR properties in SGSs to test this hypothesis for the origin of the gamma-ray excess observed in Fig.~\ref{fig:cmap}.

Particularly, we see in the right panel of Fig.~\ref{fig:cmap} that gamma-ray emission is not uniformly distributed across each of the SGS regions. This is likely a result of confinement of high-density CRs at scales smaller than the size of SGSs following our hypothesis proposed earlier. Significant correlations are expected and actually seen between these nested structures of gamma-ray emission and SNR (or candidates) populations inside the SGSs. Indeed, SNRs are believed and proven to be very efficient CR accelerators testified by their gamma-ray emission \citep{Giordano:2011ty,2013Sci...339..807A}. They are proposed to be the dominant contributor to total mechanical energy input which later drives either particle acceleration or shell expansion in the evolution of SGSs \citep{1980MNRAS.192..365M} (see also \cite{Vieu:2021yli} for more details).

\subsection{CR distribution modulated by an ensemble of SNR shocks in SGSs}\label{multiSNR}
On one hand, CR acceleration in multiple shocks and strong turbulence has been shown to be much more efficient than in the case of a single SNR \citep{1993PhyU...36.1020B,1995ICRC....3..337B,2001SSRv...99..317B}. On the other hand, when SNRs are located in a SGS, the density of CR accelerated at shock fronts are enhanced owing to the confined environment of SGSs. As a SGS expands, interstellar matter is continuously swept up and accumulates in the shell while interstellar magnetic field is compressed in the direction tangential to the shell. The super-dense shell of SGS formed in this process plays an important role in trapping the CRs \citep{Vieu:2021yli,2022MNRAS.512.1275V}, resulting in a higher CR density in the SGS interior (regarded as the region behind the super thick shell of an expanding SGS), as compared to CRs in residence of the LMC large-scale disk \citep{2016A&A...586A..71A}. These are the two important requirements for efficient production of gamma-rays.    

Different scenarios have been suggested to model the acceleration of particles in interacting shocks, such as colliding winds \citep{1990ICRC....4..105L}, supernova remnants \citep{1980ApJ...237..793B}, or an ensemble of shocks in general \citep{1988SvPhU..31...27B}. For example, the process of particle acceleration and transport in a violent bubble environment was studied comprehensively in \cite{1992MNRAS.255..269B,1993A&A...280L..27B,2001SSRv...99..317B,Parizot:2004em}. As a result of CR acceleration by multiple shocks, the particle spectra are generally hard at low energies and steepen at higher energies (~GeV) \citep{2010A&A...510A.101F,2022MNRAS.512.1275V}, although very hard spectra can be produced over a wide energy range when the acceleration sites (e.g., SBs) are highly magnetized and turbulent. 

\begin{figure}[ht]
   \centering
   \includegraphics[width=0.8\hsize]{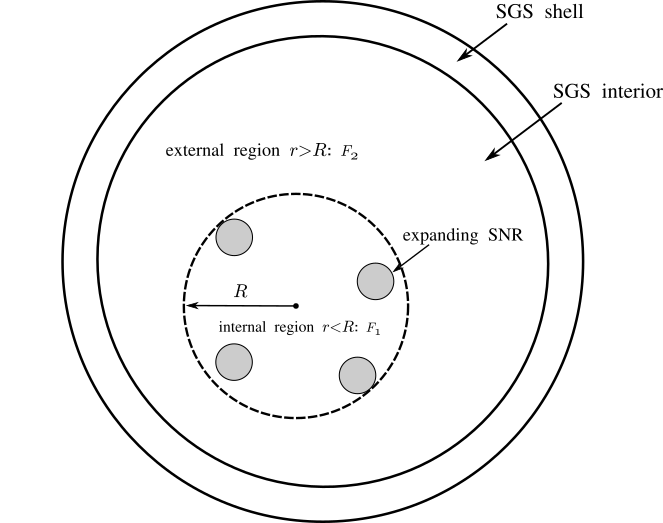}\caption{Structure view of the multi-SNR model in an SGS.}\label{fig:snr_model}
\end{figure}

For simplicity, we assume that the diffusion coefficient is sufficiently small behind each SNR shock front, such that the energetic particles are frozen into the gas flow downstream (of each SNR) and diffusion of CRs can be disregarded in the interior of each SNR envelope \citep{1988SvPhU..31...27B,2000APh....13..161K}. In regions ahead of shock fronts, the particle distribution is effectively modified by SNR shocks. Particles start to escape from the shock front of a SNR when they are accelerated to a characteristic energy $E_*$. Then particles with $E>E_*$ in between shocks simultaneously interact with several shock fronts where Fermi acceleration by large-scale turbulence occurs. Following closely the formalism developed in \cite{1988SvPhU..31...27B,2000APh....13..161K} and \cite{2001AstL...27..625B}, a steady-state equation can be constructed for the distribution of accelerated particles in between shocks when we assume for simplicity that the number of SNR shock waves remains constant in the system of a SGS\footnote{The SNR shock wave becomes subsonic and starts to dissipate once maximum shock radius is reached at $t_{\rm SNR}$, and new shock waves form as new episode of explosion begins in the SGS. The frequency of supernovae explosions expected to be larger than  $1/t_{\rm SNR}$ (see \cite{2000APh....13..161K} for detailed discussion).}. A sketch for such a setup is shown in Fig.~\ref{fig:snr_model} 
where we find for the spatial distribution of high-energy accelerated CRs with $E>E_*$ an approximating profile, expressed as \citep{2001AstL...27..625B}
\begin{equation}\label{eq:center}
    F_1(r,\,p)\sim f(p)\times\chi(r)=f(p)\times\frac{{\rm sin}(kr)}{r},
\end{equation}
where the energy distribution of accelerated particles is described by $f(p)$, and the spatial distribution follows $\chi(r)$. The parameter $k$ in $\chi(r)$ is defined as $k=\pi/2R$, and $R$ represents the size of the internal region where interactions of accelerated particles with multiple shock fronts take place. At larger radii, $r>R$, the particle distribution is determined by diffusion. The spatial profile of accelerated particles can then be described by an approximated form \citep{1988SvPhU..31...27B}, 
\begin{equation}\label{eq:upstream}
    F_2(r,\,t)\approx A\,{\rm exp}\left[-\frac{(r-R)^2}{4\kappa\,t}\right],
\end{equation}
where $A={\rm sin}(kR)/R$ is a normalization parameter determined at the internal/external region boundary. $\kappa$ is the diffusion coefficient in the upstream region for each of the SNR shocks\footnote{In our approach, we assume that all SN shocks evolve alike and there is no independence of diffusion coefficient on energy.}, and is assumed to be spatially uniform and momentum-independent in the following analysis. $t$ stands for SN shock evolution time, that is, SNR age. We can see from Eq.~\eqref{eq:upstream} that a significant number of accelerated particles with energy larger than $E_*$ is presumed to accumulate in the external region ($r>R$) with a length scale $r\sim\sqrt{\kappa\,t}$, which increases more rapidly in comparison to the size of the shock when assuming a typical relation $R_{\rm shock}(t)\sim t^{2/5}$ for the adiabatic expansion of SNR \citep{1977MoIzN....Q....S}.

\subsection{Spatial gamma-ray distribution in SGS~4 as an example}\label{subsec:lmc4}
The low-density ambient gas inside SGS\,4
follows an approximately uniform distribution according to the H$\alpha$ \citep{2003ApJS..146..407F} and HI measurements \citep{2016A&A...594A.116H}, allowing us to assume that the underlying CR distribution is well traced by the gamma-ray spatial distribution within the SGS\,4 interior. Therefore, the initial position (i.e., $r=0$) for the peak of CR density in SGS\,4 interior is set to be at the location of the highest gamma-ray emission rate, which is found to be at $(\alpha_{\rm J2000},\delta_{\rm J2000})=(83.22^\circ,-67.17^\circ)$ in the observational count map in Fig.~\ref{fig:cmap}.

By employing Eqs.~\eqref{eq:center} and \eqref{eq:upstream} of the multi-SNR model, we can construct a spatial template for gamma-ray emission from SGS\,4 . We notice that Eq.~\eqref{eq:upstream} further reduces to a form which solely depends on $-r^2/4\kappa t$ after applying the boundary condition, $F_1(R)=F_2(R)$. The spatial gamma-ray template of the SGS\,4 interior is then built using the parameters $R$ and $L$ (here we define an effective scale length\footnote{Note that this parameter should be corrected for the inclination angle of the LMC if it is to be interpreted as a physical diffusion scale length.} as $\sqrt{L}=\sqrt{4\kappa t}$). However, given the limited photon statistics in the analyzed range of $10\,\rm GeV-300\,GeV$, we first establish an exclusion mask for SGS\,4 to prevent any nearby bright sources (e.g., SGS\,5 and LMC~P3 \citep{2016A&A...586A..71A}) from contaminating the target region during the template building and fitting processes. The size of the exclusion mask is set to be equal to the radius of SGS\,4, $R_{\rm SGS\,4}=0.69^\circ$, reported in \cite{1980MNRAS.192..365M}. The spectral shape for the interior region was chosen to follow a power-law form at this stage,
\begin{equation}\label{eq:pwl}
    \left(\frac{dN}{dE}\right) = N_0\left(\frac{E}{E_0}\right)^{-\Gamma},
\end{equation}
with a photon index $\Gamma$, a normalization parameter $N_0$, and the scaling energy $E_0$.  

We perform a joint maximum-likelihood fit for the spatial and spectral parameters of the SGS\,4 interior region. This step allows us to find the best-fitting spatial parameters that lead to the largest test statistic (TS) value, defined as $TS=-2\,{\rm ln}(l_{max,0}/l_{max,1})$, where $l_{max,0}$ is the maximized likelihood value obtained with the background model (see Section~\ref{sec:fermidata}), and $l_{max,1}$ the maximized likelihood value obtained with the model of collective SNR shocks.
\begin{figure}[ht]
   \centering
   \includegraphics[width=1\hsize]{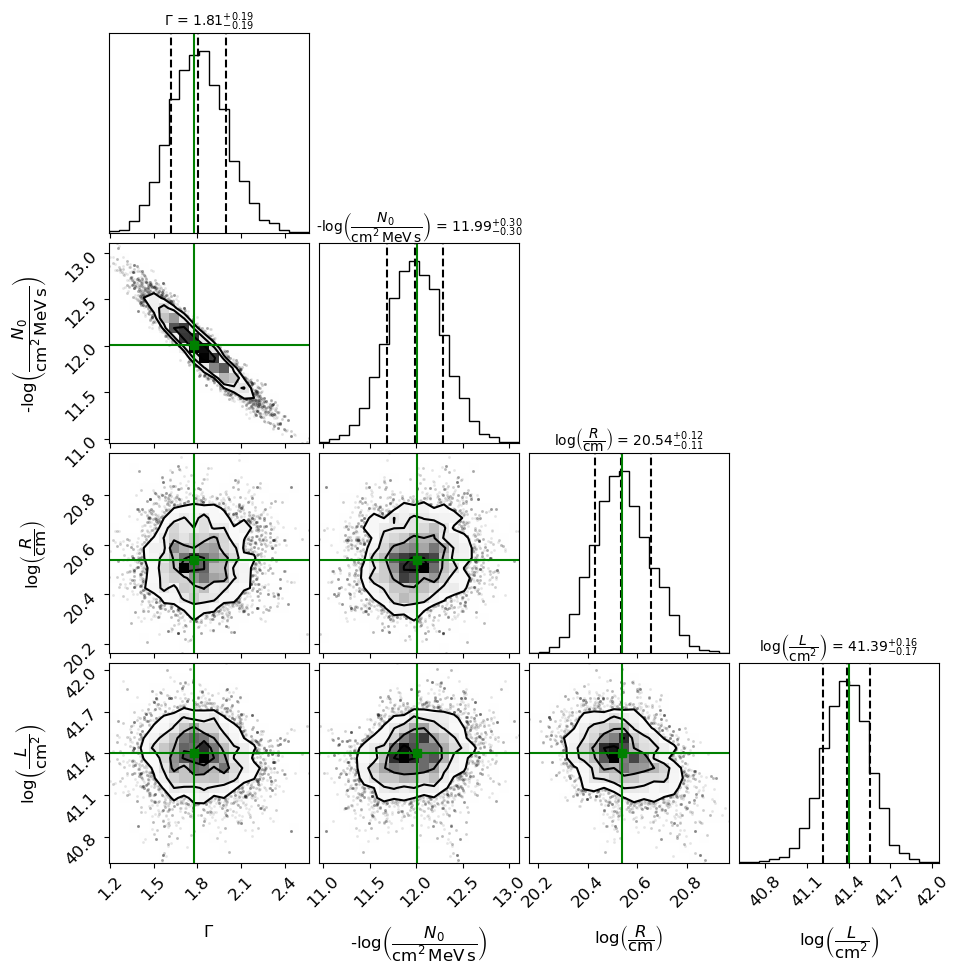}
        \caption{Corner plot of the MCMC fit to the \textit{Fermi}-LAT $10\,\rm GeV$-$300\,\rm GeV$ data in the interior of SGS\,4. The black dashed lines mark the 1$\sigma$ confidence interval for each of the fitted parameters, while the green dots and lines represent the best-fitting solutions derived from the largest TS values.}\label{fig:mcmc}
\end{figure} 
The fitting result for each parameter is shown in Fig.~\ref{fig:mcmc} with their corresponding 1$\sigma$ uncertainties. The best-fitting parameters (marked with green lines) are in general very close to the mean of samples for all 4 parameters of SGS\,4 interior. Additionally, the position for the peak CR density is re-optimized to a location of $(\alpha_{\rm J2000},\delta_{\rm J2000})=(83.01^\circ,-67.07^\circ)$ with all other spatial parameters fixed at their best fitting values. The best-fitting internal region size is $R=10^{20.54}\,\rm cm$, which translates into a value of $111.97\,\rm pc$ ($\sim0.13^\circ$). This is comparable to the maximum radius of a single SNR shock expanding in a low-density SGS environment. The best fitting result for the effective scale length is obtained as $\sqrt L=163.42\,\rm pc$. If assuming, for example, a typical value of $15\,\rm kyr$ for the age of an SNR in the stage of adiabatic expansion, we can estimate the diffusion coefficient in the external region ($r>R$) of SGS\,4 interior to be $\gtrsim1.34\times10^{29}\,\rm cm^2/s$.  

The SGS\,4 interior complex is surrounded by a thick dense shell. Faint gamma-ray emission of the shell is assessed separately from that of the interior region. In general, radial extension of the shell wall starts from the rim of a SGS interior and end at a distance where the gas surface density has fallen to its first minimum \citep{2013ApJ...763...56D}. In the case of SGS\,4, we assume for simplicity that the thickness of SGS\,4 shell is equal to 20\% of the interior region radius, and we model the spatial distribution of the shell emission using \textit{ShellSpatialModel}, which is conveniently implemented in the \texttt{Gammapy}\footnote{\url{https://gammapy.org/}} Python package \citep{2023A&A...678A.157D}. As for the shell spectrum, we take the same power-law form as used in the case of the interior region, thus only two free parameters ($N_0$ and $\Gamma$) are used to generate the shell template.

A joint fit is then performed for the spectral parameters of the SGS\,4 interior and shell regions with all spatial parameters fixed at their best values. In this way, we obtain a maximized TS value of 132.04 for the optimized SGS\,4 template in between $10\,\rm GeV$ and $300\,\rm GeV$.

\subsection{Other SGSs in the LMC}\label{subsec:othersgs}
Strong signals with detection above $3\sigma$ (TS $\sim$ 9) are consistently seen in regions of SGS\,2, 3, 4, 5, and 7 in the middle and right panels of  Fig.~\ref{fig:cmap} (see also Fig.~\ref{fig:resi_10gev} in Appendix~\ref{app:resi_10gev}). In order to have a better description for the gamma-ray distribution in the interior of these SGSs, they are then modeled with a template that is built using the same approach as exemplified in detail in the case of SGS\,4 (see  Subsection~\ref{subsec:lmc4}).
We show in Table~\ref{table:1} our best-fitting spatial parameters obtained for the regions of SGS\,2, 3, 4, 5 and 7 in the energy range of 10\,GeV-300\,GeV. These spatial parameters are then used for the creation of SGS templates in these regions. 

We obtain best-fitting values of $R$ close to $\sim300\,\rm pc$ for SGS\,2 and 3, indicating a sizable region for efficient particle acceleration by multiple shocks. On the other hand, the small value of $R\sim100\,\rm pc$, seen in SGS\,4 and 7, for instance, can be regarded as an upper bound on the size of the internal region, since it is at the limit of the LAT PSF for energies above 10\,GeV. This small size may suggest that the SNR shock fronts within these SGSs are either sparse or widely distributed from each other, making the formation of a turbulent, large internal region for efficient multi-shock acceleration less likely. Consequently, a substantial fraction of accelerated particles would accumulate at large radii $r>R$ where diffusion governs the particle spatial distribution. This scenario is especially supported in the case of SGS\,4, for which we find a large value of $\sqrt{L}$. In contrast, for most SGSs listed in Table~\ref{table:1}, the effective scale lengths are generally small, $\sim10\,\rm pc$, implying that diffuse emission at large radii is not significantly detected. Thus, the excess emission in these SGSs is primarily attributed to the internal region (with a radial extent of $0.2^\circ\sim0.3^\circ$), which, could also be interpreted as a result of a collection of faint, unresolved sources.

\begin{table*}
\caption{Best-fitting spatial parameters for the interior region of significantly detected SGSs in the 10-300\,GeV band.}              % title of Table
\label{table:1}      % is used to refer this table in the text
\centering
\renewcommand{\arraystretch}{1.5}% used for centering table
\begin{tabular}{c c c c c c c c c}         % centered columns (4 columns)
\hline             % inserts double horizontal lines
& \multicolumn{5}{c}{SGS interior} & \multicolumn{2}{c}{SGS shell} &\\
\cmidrule(r){2-6}\cmidrule(r){7-8}
Region & $\dfrac{R}{[\rm pc]}$ (1$\sigma$) & $\dfrac{R}{[\rm pc]}$ (best) & $\dfrac{\sqrt{L}}{[\rm pc]}$ (1$\sigma$) & $\dfrac{\sqrt{L}}{[\rm pc]}$ (best) &  $\dfrac{(\alpha,\,\beta)_{\rm J2000}\,\,{\rm at}\,\,r=0}{[\rm deg]}$  & $\dfrac{R_{\rm SGS}}{[\rm pc]}$ & $\dfrac{R_{\rm shell}}{[\rm pc]}$ & TS\\ 
\midrule
    SGS\,2 & $259.24^{+20.26}_{-15.93}$ &254.99 &$12.52^{+5.04}_{-4.67}$ &12.53 &$(84.53\pm0.05,-69.22\pm0.02)$ &450  &540 &506.91 \\ \hline   
    SGS\,3 & $282.33^{+37.38}_{-31.28}$ &274.83 &$13.62^{+5.69}_{-5.29}$ &12.95 &$(82.56\pm0.06,-69.09\pm0.01)$ &500  &600 &484.61 \\ \hline
    SGS\,4 & $111.47^{+31.14}_{-28.70}$ &111.97 &$159.77^{+30.17}_{-31.67}$ & 163.42  & $(83.03\pm0.08,-67.07\pm0.06)$ &600  &720 &132.01 \\ \hline
    SGS\,5 & $148.60^{+46.36}_{-48.79}$ &142.15 &$30.90^{+22.07}_{-28.70}$ & 54.96    &$(81.23\pm0.05,-66.12\pm0.07)$ &400  &480 &42.25 \\ \hline
    SGS\,7 & $100.37^{+20.65}_{-19.53}$ &100.01 &$9.83^{+3.08}_{-3.70}$ &10.80    &$(74.02\pm0.08,-69.31\pm0.03)$ &400  &480 &37.97 \\
\hline                                             %inserts single line
\end{tabular}
\tablefoot{$R(L)$ (1$\sigma$) represents the mean value and propagated uncertainty obtained with the MCMC method, while $R(L)$ (best) indicates the best-fitting parameter. The maximum TS value gives the total detection significance of the interior and shell components in each SGS region. $R_{\rm SGS}$ is the radius of the SGS interior, which is also the distance of inner edge of the SGS shell to its center. The $R_{\rm SGS}$ values were taken from \cite{1980MNRAS.192..365M}. $R_{\rm shell}$ is the distance of the outer boundary of the SGS shell to its center, which is assumed to be 20\% larger than $R_{\rm SGS}$ for simplicity.}
\end{table*}

Additionally, very low detection TS values are observed in the rest of the regions (i.e., SGS\,1, 6, 8, and 9; see Fig.~\ref{fig:resi_10gev}), which might imply fewer SNRs inside these SGSs or that the shocks are too far apart from each other, leading to a low-efficiency of particle acceleration, similar to the earlier discussion of SGSs that have rather small values of $R$. Low photon statistics could induce great uncertainties when fitting the sensitive parameters of $R$ and $L$ in these regions. Consequently, we choose a simplified 2D Gaussian spatial template for modeling gamma-ray distributions in these SGSs, where the Gaussian kernel size is equal to the radius of SGS given in \cite{1980MNRAS.192..365M}. The spectra for these SGSs (SGS\,1, 6, 8, and 9) are extracted under assumption of a simple power-law model.     

\section{Extended energy band: 300~MeV-300~GeV}\label{sec:full_band}
In the lower energy band (<$10\,\rm GeV$), the large-scale diffusion emission of the LMC becomes prominent and needs to be accounted for in the source models. As was pointed out in previous studies, the spatially extended LMC is dominated by diffuse gamma-ray emission produced in interactions between CRs and the ISM (and/or the ISRF). Using radiation field and tracers of matter, we can determine the relative contributions of various mechanisms (e.g., leptonic, hadronic) to the large-scale gamma-ray signal from the LMC. As shown by results in \cite{2010A&A...512A...7A,2015ApJ...808...44F,2016A&A...586A..71A}, strong evidence supports that the spatial profile of H$\alpha$ yields the best fit for mimicking the gamma-ray distribution produced by hadronic process in the LMC, compared to other common tracers such as HI or CO (or $\rm H_2$). However, H$\alpha$ emission does not trace thermal protons in a way as for the case of HI or $\rm H_2$. Thus we consider a square root of H$\alpha$ to be our gas template as a proxy physical tracer for ionized gas, following the same argument stated in \cite{2015ApJ...808...44F}. On the other hand, the ISRF emits gamma-rays produced in leptonic processes, in which CMB, starlight and dust-related emissions are the main components. CMB can be absorbed in the spectral normalization parameter since it is largely isotropic, and starlight was shown to provide a poor fit to the spatial gamma-ray distribution in the LMC (see Appendix~\ref{app:others}). Instead, dust emission, of all ISRF tracers, was proven to provide the best fit to the large-scale gamma-ray signal of the LMC \citep{2015ApJ...808...44F}. We use 160$\mu\rm m$ emission map as a tracer for the contribution from dust, based on the global SED of the LMC \citep{2010A&A...519A..67I,2013AJ....146...62M,2015ApJ...808...44F}. We take the $\sqrt{\rm H\alpha}$ and 160$\mu\rm m$ templates as our fiducial tracers to account for contributions of the hadronic and leptonic processes to the LMC gamma-ray signal at large scale\footnote{Various ISM, ISRF, or combined ISM/ISRF tracers have been thoroughly discussed in \cite{2015ApJ...808...44F}. The combination of $\sqrt{\rm H\alpha}$ and 160$\mu\rm m$ was shown to provide the best Log-likelihood values in our analysis, in comparison to ISM/ISRF alone or other combinations of available ISM/ISRF tracers.}, although it is very likely they do not fully account for all gas/ISRF-induced gamma-ray emission and it cannot be excluded that a small fraction of the gamma‑rays attributed to the SGS originates from large‑scale diffuse emission (see also Fig.~\ref{fig:new_sig}). The tracer maps of ionized gas H$\alpha$ and dust-related emission 160$\mu\rm m$ are extracted as a circular region of $5^\circ$ in radius around on our target ($(\alpha_{\rm J2000},\delta_{\rm J2000})=(80.00^\circ,-68.75^\circ)$) from the full-sky composite H$\alpha$ map \citep{2003ApJS..146..407F} and AKARI far-infrared all-sky survey \citep{2015PASJ...67...50D}, respectively. Foreground Galactic contribution to the observed emission for both maps is estimated from the off-source intensity in the nearby area outside the ROI region, and subtracted from the circular region.

\begin{figure}[ht]
   \centering
   \includegraphics[width=0.8\hsize]{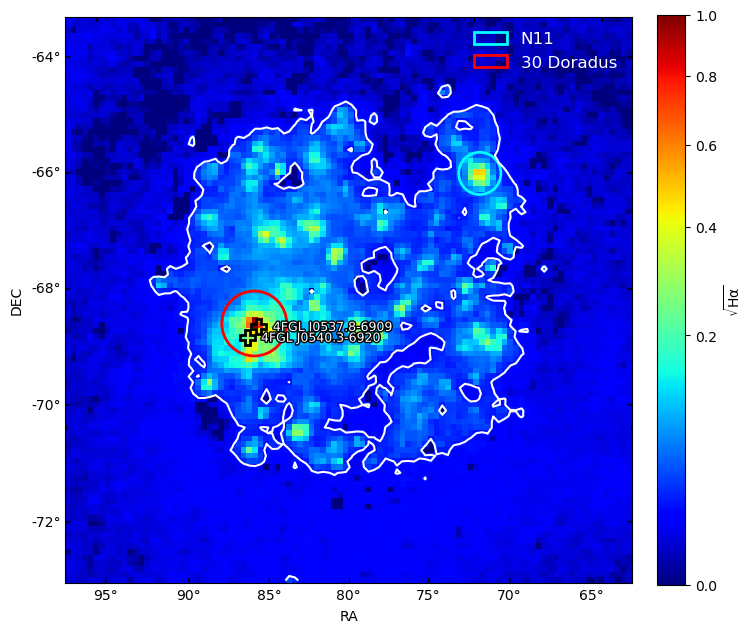}
   \includegraphics[width=0.8\hsize]{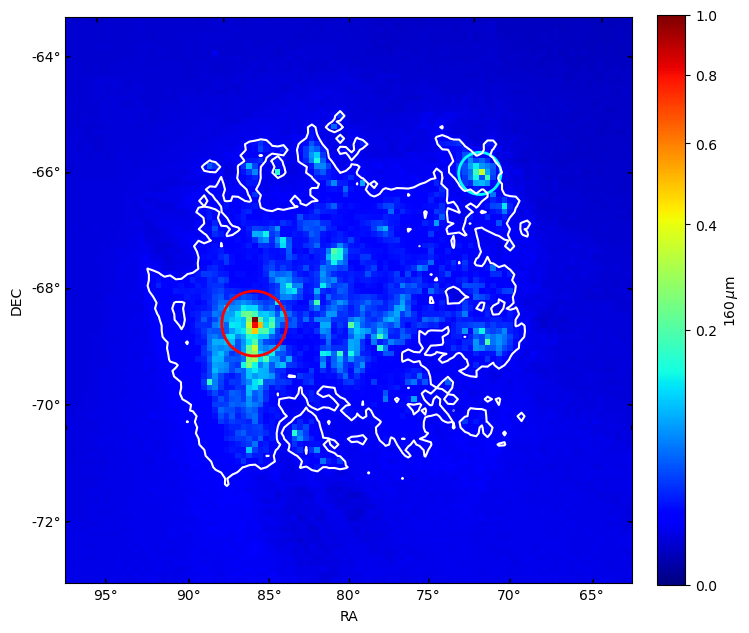}
      \caption{\textit{Upper}: Square root of the composite H$\alpha$ emission map \citep{2003ApJS..146..407F} centered on the LMC. \textit{Lower}: $160\mu\rm m$ \citep{2015PASJ...67...50D} radiation field centered on the LMC. Both maps are normalized so that their maximum emission values equal one (dimensionless). White contours indicate about 3\% and 1\% levels of the maximum emission intensities in $\sqrt{\rm H\alpha}$ and $160\mu\rm m$, respectively. Star-forming regions of 30 Doradus is marked with a red circle, and N11 with a cyan circle in both maps. Point sources detected by \textit{Fermi}-LAT within the two star-forming regions are denoted with black-outlined crosses in the upper panel.}\label{fig:ism/isrf}
\end{figure}

Additionally, it is important to notice that the $\sqrt{\rm H\alpha}$ and $160\mu\rm m$ maps peak at two locations which corresponds to the two largest star-forming regions in the LMC: 30 Doradus and N11 (see Fig.~\ref{fig:ism/isrf}). Such feature of the tracer templates would yield gamma-ray sources within or in the vicinity of 30 Doradus and N11 regions to negligible detection TS values, leading to a biased modeling template for the isolated sources and global examination of gamma-ray flux at large scale.
For example, \cite{2016A&A...586A..71A} showed that using an ionized-gas template produces an apparent enhancement of gamma-ray emission toward the 30 Doradus region, owing to the peak feature of the gas template in that region. However, the authors noted that the bright gamma-ray emission from 30 Doradus is primarily caused by the presence of pulsar PSR J0540-6919. In this way, the inferred CR density from the template fitting for the large-scale of the LMC would not accurately represent the true property of CR population distributed across LMC\footnote{N11 and 30 Doradus are extreme star-forming regions packed with very young massive stars ($<5\,\rm Myr$), but only a few supernovae exploded in the recent past could have injected or accelerated CRs significantly (see also \cite{2023MNRAS.523.5353A}). Therefore, these bright regions are regarded as statistical outliers and could lead to incorrect assumption on the CR-induced gamma-ray emission.}. 
Therefore, contributions from the two regions where the most significant peaks appear are removed from the two templates shown in Fig.~\ref{fig:ism/isrf}, such that a better assessment on the large-scale component of the gamma-ray emission can be achieved in the following analysis. The corresponding spectral shapes for the two templates of $\sqrt{\rm H\alpha}$ and $160\mu\rm m$ are found to prefer a log parabola (LogP) model against a simple power law (PL), 
\begin{equation}\label{eq:lp}
\left(\frac{dN}{dE}\right) = N_0\left(\frac{E}{E_0}\right)^{-(\Gamma+\beta\,{\rm ln}(E/E_0))},
\end{equation}
where $\beta$ is the curvature parameter. This is consistent with results reported in \cite{2016A&A...586A..71A}. Finally, these two spatial templates, each with their respective spectral parameters, are used to model the gamma-ray flux assuming uniformly distributed CR particles in the LMC (see however Appendix~\ref{app:others} for a comparison of other assumed CR distributions at large scales in the LMC).  

Then, we fix the spatial template parameters of all SGS regions at their best values obtained in Subsections~\ref{subsec:lmc4} and \ref{subsec:othersgs}, and employ them in the analysis of the full energy band of $300\,\rm MeV$-$300\,\rm GeV$. Together with the large-scale tracer templates of $\sqrt{\rm H\alpha}$ and  $160\mu\rm m$, we perform a joint likelihood analysis. In the final step, we add 7 previously detected sources from the \textit{Fermi}-LAT 4FGL-DR3 source catalog (4FGL~J0511.4-6804, 4FGL J0524.8-6938, 4FGL J0535.2-6736, 4FGL J0537.8-6909, 4FGL J0540.3-6920, 4FGL J0517.9-6930c, 4FGL J0535.7-6604c; see, e.g., Fig.~\ref{fig:cmap}), which are located within the boundary of the LMC, to our source model of the ROI. In addition, we note that the presence of SGS shells in the $\sqrt{\rm H\alpha}$ template could, in principle, introduce systematic biases in the flux estimates derived for the large-scale $\sqrt{\rm H\alpha}$ template and SGS regions during the joint likelihood fitting procedure. For this, we masked the SGS regions in the $\sqrt{\rm H\alpha}$ template and repeated the joint likelihood analysis. We found that the resulting SEDs of the large-scale and SGS components were consistent with those obtained when the SGS regions were not masked in the $\sqrt{\rm H\alpha}$ template, that is, the deviations between the two cases remain within $1\sigma$ uncertainties.

In Table~\ref{tab:joint} we list the best results for a joint fit of the three major components in the analyzed ROI: 4FGL-DR3 isolated sources (10 background sources outside the LMC boundary, and 7 sources from within the LMC boundary; component A), small-scale regions (SGSs; component B), and the large-scale diffusion region across the LMC (component C). In addition, we also provide in models 3 and 4 the joint fit results, of which all 9 SGSs are modeled with a simplified Gaussian template, as comparisons to models 1 and 2. 
\begin{table}[ht]
\caption{Maximized log-likelihood results with a joint fitting of components A, B, and C.}             % title of Table
\label{tab:joint}    
\renewcommand{\arraystretch}{1.7}
\centering
\begin{tabular}{ c|c|c|c } 
\hline
model & B & C & $l_{max,\,i}$ \\
\hline
\makecell{Background\\($n_{dof}=25$)} & - & - & 68044.9 \\
\hline
\makecell{4FGL-DR3 \\catalog\\($n_{dof}=55$)}  & - & - &  87406.8\\
\hline
\makecell{model 1\\($n_{dof}=75$)}  & \makecell{SGS\,2-5 and 7: \\multi-SNR model;\\ SGS\,1, 6, 8, and 9:\\Gaussian template} & \makecell{with \\30 Doradus\\and N11} & 87541.8 \\
\hline
\makecell{model 2\\($n_{dof}=75$)}  & \makecell{SGS\,2-5 and 7: \\multi-SNR model;\\ SGS\,1, 6, 8, and 9:\\Gaussian template} & \makecell{without \\30 Doradus\\or N11} & 87555.0 \\
\hline
\makecell{model 3\\($n_{dof}=65$)}  & \makecell{SGS\,1-9: \\Gaussian template} & \makecell{with \\30 Doradus\\and N11} & 87459.3 \\
\hline
\makecell{model 4\\($n_{dof}=65$)}  & \makecell{SGS\,1-9: \\Gaussian template} & \makecell{without \\30 Doradus\\or N11} &  87423.4\\
\hline
\end{tabular}
\tablefoot{Component A is the same for models 1-4 listed below, whereas components B and C have different spatial modeling options. The power-law spectral shape was taken for component B of models 1-4, while a LogP form was chosen for the large-scale templates in component C. $n_{dof}$ stands for the total degrees of freedom in the respective models.}
\end{table}
The maximum likelihood value is achieved with model 2, and it shows an overall increasing TS value of $\sim200$ at the cost of additional 10 degrees of freedom compared to models 3 and 4, indicating an non-Gaussian gamma-ray distribution in SGSs interior. 
In comparison, a best-fitting log-likelihood value (see Table~\ref{tab:joint}) is also obtained when we employ a ROI source model including all point-like and extended sources stated in the 4FGL-DR3 catalog.

\section{Spectral analysis results}\label{analysis}
\begin{figure}[ht]
   \centering
   \includegraphics[width=0.8\hsize]{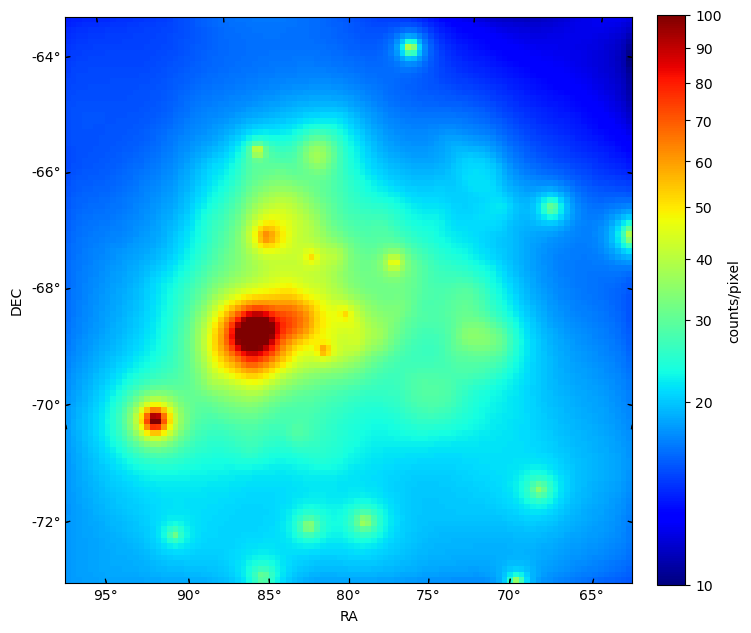}
   \includegraphics[width=0.8\hsize]{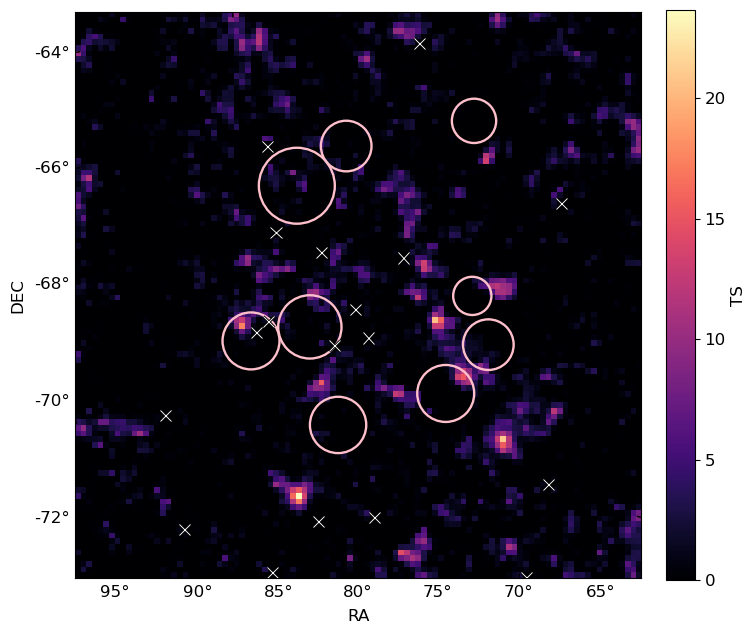}\caption{Predicted count map (\textit{top} panel) of model 2 with relevant point-like sources and its corresponding residual TS map (\textit{bottom} panel), which has a maximum value of $\rm TS=23.66$. White crosses in the bottom panel indicate positions of all relevant point-like sources within the ROI.}\label{fig:final_resimap}
\end{figure}

In this section, we present the spatially integrated gamma-ray spectra emitted from small-scale SGS regions and the large-scale region of the LMC with the observational data presented in the left panel of Fig.~\ref{fig:cmap}, where 18 logarithmic energy bins are evenly distributed between 300\,MeV and 300\,GeV. When applying our preferred model (model 2 in Table~\ref{tab:joint}) to estimate fluxes from different regions of the LMC, we inspected the residual TS map of model 2 and identified two significant residuals with detections above $5\sigma$.
We therefore added two point sources (PS1 and PS2) to account for these excesses to ensure no contamination from them when estimating fluxes of individual components in model 2. PS1 was identified at a position of $(\alpha,\delta)_{\rm J2000}=(81.81^\circ,-68.03^\circ)$ with a 95\% confidence radius of $0.07^\circ$ and a TS value of 43.11; PS2 at $(80.15^\circ,-68.97^\circ)$ with a 95\% confidence radius of $0.10^\circ$ and $\rm TS=25.82$. No gamma-ray source counterparts in SIMBAD are found for PS1 and PS2, nor are they listed in the \textit{Fermi} 4FGL catalog. Their spectra can be described by a PL with photon indices 2.55 and 2.51, respectively. A re-fitted model count map for model 2 including PS1 and PS2 is shown in the upper panel of Fig.~\ref{fig:final_resimap}, while the corresponding residual TS map is shown in the lower panel (see also a residual significance map for the re-fitted model in Appendix~\ref{app:new_sig}). We also show a decomposition of components in model 2 in Appendix~\ref{app:compb_c}. 

Maximized test statistic values for the large- and small-scale components in the re-fitted ROI model are listed in Table~\ref{tab:sedparas}. The large-scale components, $\sqrt{\rm H\alpha}$ and $160\mu\mathrm{m}$, yield a summed TS of 2696.47. Among the small-scale components, SGS\,2-8 is significantly detected. The most significant emission is detected toward the region of SGS\,2, with the dominant contribution arising from its interior ($\rm TS=1691.38$) and only a small fraction from its shell ($\rm TS_{shell}=121.34$). A similar interior-dominated emission pattern is seen in SGS\,3–5, although TS values of their shell components are much smaller and fall below the detection threshold. In contrast, emission from SGS\,7 is detected primarily in its shell ($\rm TS_{shell}=47.48$), with no significant detection from the interior. This may indicate that the present SNR population in SGS\,7 is concentrated near the shell rather than in its interior. 

\begin{figure*}[ht]
   \centering
   \includegraphics[width=0.246\hsize]{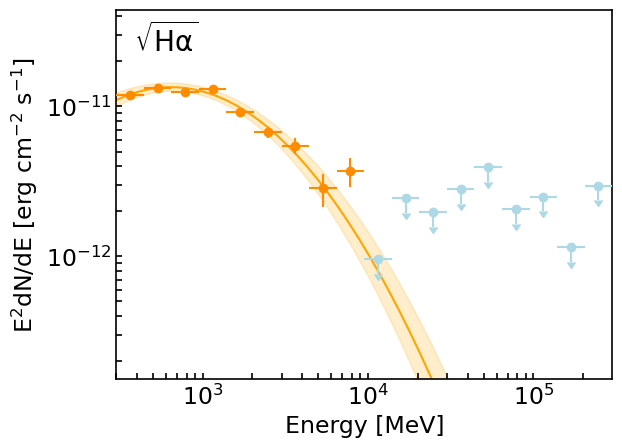}
   \includegraphics[width=0.246\hsize]{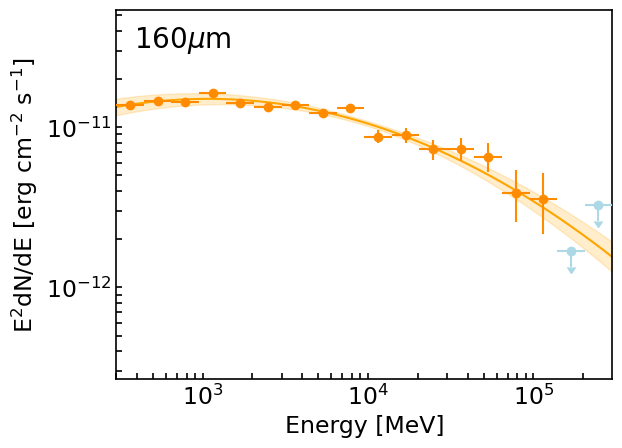}
   \includegraphics[width=0.246\hsize]{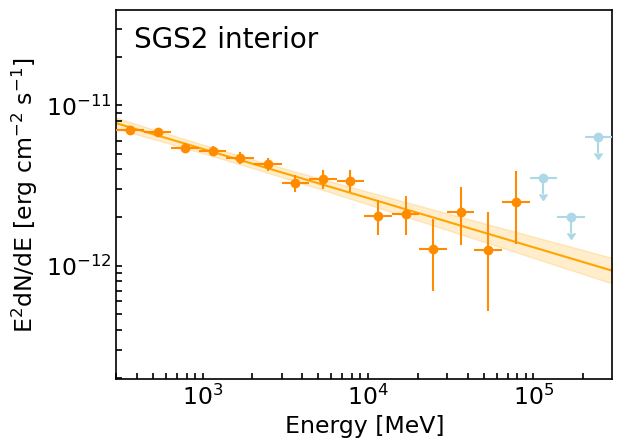}
   \includegraphics[width=0.246\hsize]{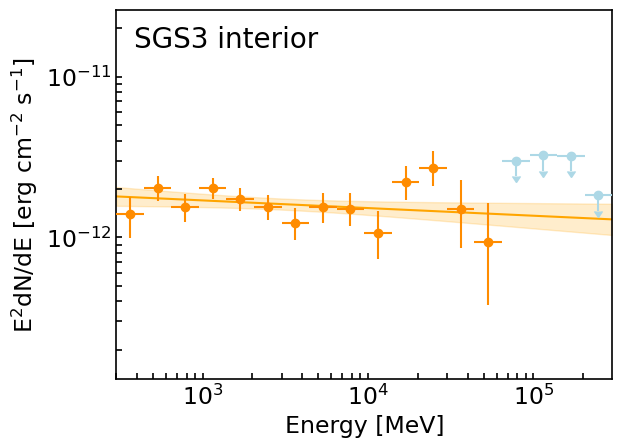}
   \includegraphics[width=0.246\hsize]{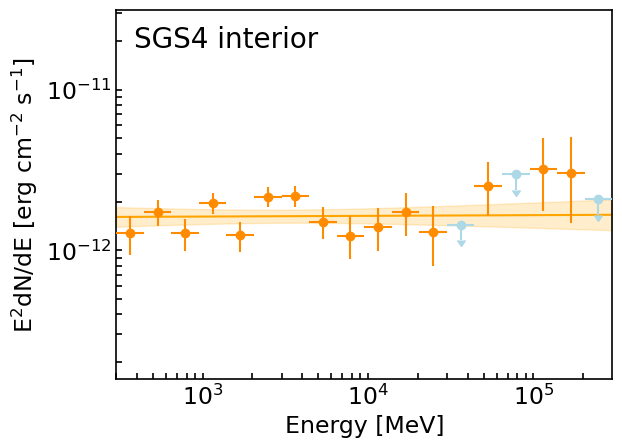}
   \includegraphics[width=0.246\hsize]{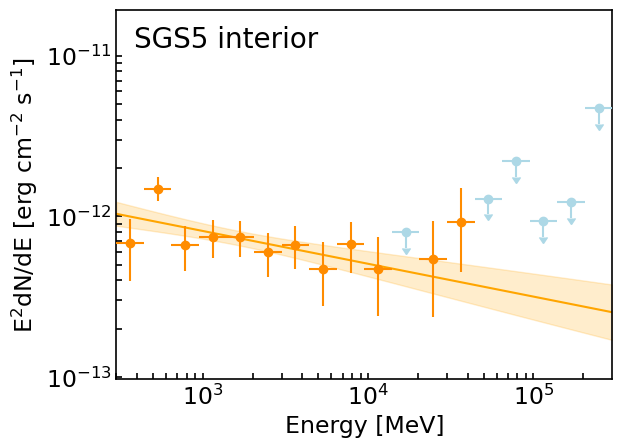}
   \includegraphics[width=0.246\hsize]{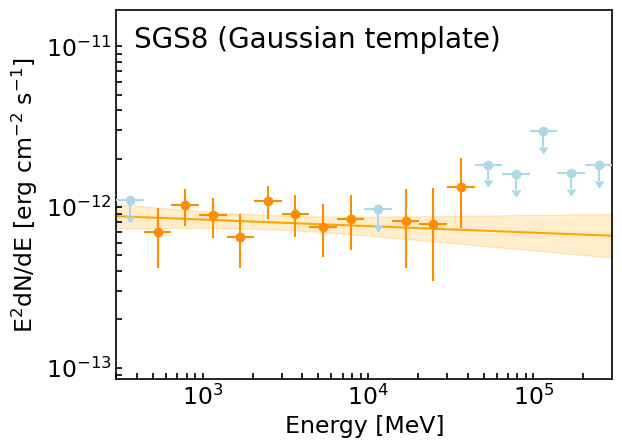}
   \includegraphics[width=0.246\hsize]{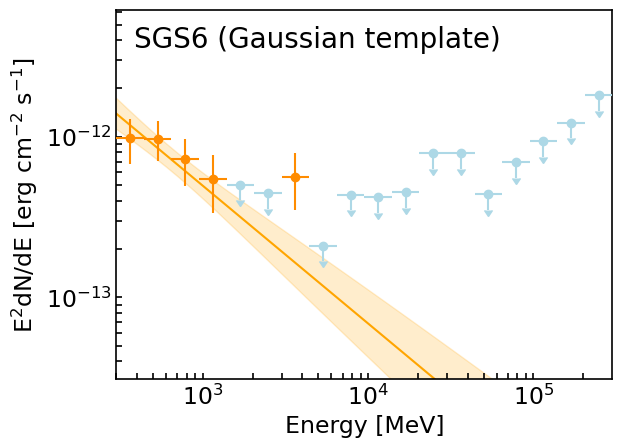}
   \includegraphics[width=0.246\hsize]{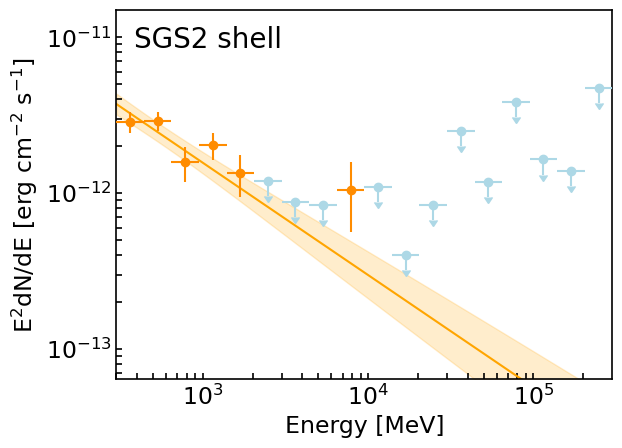}
   \includegraphics[width=0.246\hsize]{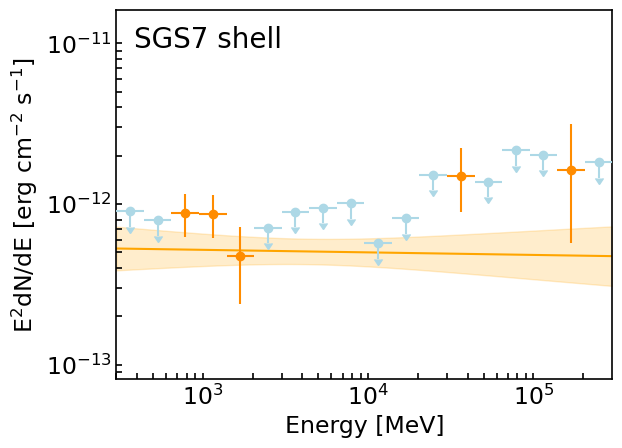}
   \caption{Spectra based on templates of $\sqrt{\rm H\alpha}$ and $160\mu\rm m$ and significantly detected SGSs ( SGS interiors or SGS shells). Spectral data points are obtained by fits in individual bins, together with the best-fit spectral model from the binned maximized likelihood analysis over the full energy band. Orange error bars represent a $1\sigma$ flux uncertainty on the data points, while the solid lines and shaded area stand for the best-fit spectral model and its $1\sigma$ uncertainty respectively. The light blue upper limit bars correspond to a 95\% confidence level.}\label{fig:seds}
\end{figure*}

The gamma-ray fluxes from different spatially integrated regions are estimated using respective components of model 2 shown in Table~\ref{tab:joint}. As a result, we show in Fig.~\ref{fig:seds} the SEDs for various SGSs (interior) and the large-scale diffusion components of model 2. Also, we show the SEDs of two significantly detected shell components (i.e., SGS\,2 shell and SGS\,7 shell).  
Spectral parameters of nearby sources within $3^\circ$ of the target source are left free and re-fitted for each case shown in Fig.~\ref{fig:seds}. The corresponding fitting results are shown in Table~\ref{tab:sedparas}. 

\begin{table*}[ht]
\caption{Best-fitting spectral parameters of the large-scale components and SGSs (interior) from the 300\,MeV-300\,GeV analysis.}             % title of Table
\label{tab:sedparas}    
\renewcommand{\arraystretch}{1.4}
\centering
\begin{tabular}{@{\extracolsep{30pt}}cccccc}
%\begin{tabular}{ccccc}
\hline\hline
 source &$E_{\rm flux,>1GeV}$ & $N_0$ & $\Gamma$ & $\beta$ & TS \\
\hline
$\sqrt{\rm H\alpha}$ & 1.46$\pm$0.15 &0.79$\pm$0.05& 2.30$\pm$0.05& 0.34$\pm$0.06 &689.56\\
$160\mu\rm m$        & 4.92$\pm$0.62 &0.94$\pm$0.08& 1.99$\pm$0.04& 0.07$\pm$0.01 &2006.91\\
SGS\,2 interior      & 1.44$\pm$0.14 &0.33$\pm$0.02& 2.31$\pm$0.04&-  &1691.38\\
SGS\,3 interior      & 0.85$\pm$0.15 &0.11$\pm$0.01& 2.05$\pm$0.05&-  &397.09\\
SGS\,4 interior      & 0.94$\pm$0.17 &0.10$\pm$0.01& 2.00$\pm$0.05&-  &385.34\\
SGS\,5 interior      & 0.27$\pm$0.07 &0.05$\pm$0.01& 2.20$\pm$0.09&-  &142.33\\
SGS\,8               & 0.42$\pm$0.10 &0.05$\pm$0.01& 2.04$\pm$0.07&- &118.15\\
SGS\,6               & 0.06$\pm$0.02 &0.03$\pm$0.01& 2.86$\pm$0.22&- &39.92\\
SGS\,9               & $<$0.02 &-&-      &- &11.10\\
SGS\,7 interior      & $<$0.10 &-& -      &- &6.81\\
SGS\,1               & $<$0.0006&- &- &-&0.05\\
\hline
SGS\,2 shell         & $0.21\pm0.05$& $0.10\pm0.02$& $2.72\pm0.13$&- &121.34 \\
SGS\,7 shell         & $0.28\pm0.12$& $0.03\pm0.01$& $2.02\pm0.12$& -&47.48 \\ 
SGS\,3 shell         & $<$0.57& -& -& -&23.39 \\
SGS\,4 shell         & $<$0.42& -& -& -&15.48 \\ 
SGS\,5 shell         & $<$0.24& -& -& -&2.01 \\
\hline
\end{tabular}
\tablefoot{$E_{\rm flux,>1GeV}$ is the integrated energy flux between 1\,GeV and 300\,GeV in units of $10^{-11}\,\rm erg\,cm^{-2}\,s^{-1}$. $N_0$ is the differential photon flux evaluated at 1\,GeV in units of $10^{-11}\,\rm MeV^{-1}\,cm^{-2}\,s^{-1}$. The listed errors are $1\sigma$ uncertainties. A 95\% confidence level is shown for energy fluxes of certain SGSs where only an upper limit exists for most of the spectral bins. Additionally, we list the maximized TS values for the large- and small-scale components.}
\end{table*}

As it can be seen from the integrated energy fluxes (over 1-300\,GeV band) listed in the Table~\ref{tab:sedparas}, components of $\sqrt{\rm H\alpha}$ and $160\mu\rm m$ contribute a total of $6.38\times10^{-11}\,\rm erg\,cm^{-2}\,s^{-1}$ to diffuse emission of the LMC. In contrast, $4.48\times10^{-11}\,\rm erg\,cm^{-2}\,s^{-1}$ is the sum of all predicted energy fluxes of significantly detected SGSs (i.e., SGS\,2-5 interior, SGS\,2 shell, SGS\,7 shell, SGS\,6, SGS\,8), which is about 70\% of the energy emission from large scale of the LMC, and about 41\% of the total diffuse emission of the entire LMC. An upper limit for cumulative emissions from shell components of SGS\,2-5, 7 is estimated to be $1.83\times10^{-11}\,\rm erg\,cm^{-2}\,s^{-1}$ at a 95\% confidence level.

In general, the detected gamma-ray emission for just a few SGSs is in good agreement with the findings of nonthermal synchrotron emission at GHz bands by rough estimates (see, e.g., Figure~13 in \cite{2022MNRAS.510...11H}). For example, the most significant emission in the gamma-ray and radio bands seems to coincide with SGS\,2. This leads to an important implication that contribution of leptonic processes to the emission of gamma-rays in SGS\,2 is nontrivial and may contribute significantly, for example, by the young massive star cluster R136 \citep{2024ApJ...970L..21A} and SB 30 Dor C within SGS\,2 (see also discussion in Section~\ref{intro}). In contrast, faint SGS regions measured in radio analysis may indicate an insignificant contribution of a leptonic component to the gamma-ray emission.

On the other hand, we can view the significance of hadronic contribution to the total emission by comparing the CR proton energy density in individual SGS and then to the large-scale emission component. The CR particles are expected to have a harder spectrum in SGS regions than that of the large-scale region. This conjecture is supported when we calculate the CR proton energy density from the integrated energy flux listed in Table~\ref{tab:sedparas} using the expression given in \cite{2022A&A...659A.105S},
\begin{equation}
    \omega_{\rm CRp,\,>10{\rm GeV}} = \frac{L_{\gamma,\,>1{\rm GeV}}}{\kappa_\pi\,\sigma_{p-p}\,c\,\eta_N\,N_{\rm H}\,S},
\end{equation}
where the gamma-ray luminosity is calculated as $L_{\gamma,\,>1{\rm GeV}}=4\pi d^2E_{\rm flux,\,>1GeV}$, and $d$ is the distance to LMC. For protons above $1\,\rm GeV$, the total inelastic p-p cross-section $\sigma_{p-p}$ is approximately $30\,\rm mb$, and $\kappa_\pi\approx0.45$ for the fraction of kinetic energy of high energy protons transferred in $\pi^0$ production \citep{2018A&A...615A.108Y}. $\eta_N\approx1.5$ accounts for contribution from heavier nuclei \citep{2016Natur.531..476H}. $N_{\rm H}$ is the ambient hydrogen column density in the observed area $S$. This way, by substituting the respective values of $E_{\rm flux,\,>1GeV}$ in Table~\ref{tab:sedparas} and $N_{\rm HI}$ of each SGS in \cite{2001ApJS..136...99P} as an approximating value for $N_{\rm H}$\footnote{The ionized hydrogen column density of the LMC can be estimated from the H$\alpha$ map \citep{2003ApJS..146..407F} using a conversion factor $N_{\rm HII}=1.37\times10^{18}\times I_{\rm H\alpha}$ \citep{2010A&A...512A...7A}. $N_{\rm H}$ would increase $<10\%$ for all SGS regions when including the ionized hydrogen, and $<4\%$ for the LMC disk. It is thus neglected in our analysis.}, we compute the CR proton energy densities in SGS\,2-5 to be $0.20\,\rm eV/cm^3$,  $0.23\,\rm eV/cm^3$, $0.39\,\rm eV/cm^3$,
and $0.15\,\rm eV/cm^3$, respectively. In contrast, $\omega_{\rm CRp}$ for the large scale LMC disk arrives at a value of $0.04\,\rm eV/cm^3$ using a mean column density of $N_{\rm H}=1.4\times10^{21}$ H-atom$/\rm{cm^{2}}$ \citep{2015ApJ...808...44F} and the total integrated energy fluxes of $\sqrt{\rm H\alpha}$ and $160\mu\rm m$. This is a factor of about 3-10 smaller as compared to the local $\omega_{\rm CRp}$ calculated in SGS regions, which is consistent with the findings on CR density in small-scale extended emission regions in \cite{2016A&A...586A..71A}.

The spectral origin of SGS interior is studied in detail to determine its dominant emission mechanism. We again use SGS\,4 as an example to show our approach. The measured spectrum of interior from SGS\,4 is explored using a radiation model implemented in the \texttt{Naima}\footnote{See the references therein \url{https://naima.readthedocs.io/en/latest/index.html} for related parameters input that are not mentioned here.} package \citep{naima}, which allows to assess for emissions from processes of neutral pion decay of hadronic collisions, electron bremsstrahlung, and IC scattering from a parent population of CRs. Electron bremsstrahlung is formulated based on the electron-ion interaction model provided in \texttt{Naima}. The property of ISM ions in the LMC is characterized by an average density of $n_i=0.012\,\rm cm^{-3}$  \citep{2002A&A...392..103S,2006A&A...447..991C} and metallicity of $Z^2=1.2$ \citep{2002A&A...392..103S}. We considered a PL and PL with an exponential cutoff 
for the distribution of electron(proton) spectrum, $dN_e/dE_{e}$. The PL formula follows the same form given in Eqs.~\eqref{eq:pwl}, while the generalized PL with an exponential cutoff (PLExp) has the form of
\begin{equation}\label{eq:plexp}
    \left(\frac{dN_e}{dE_e}\right) = N_{e,0}\left(\frac{E}{E_0}\right)^{-\Gamma}\,{\rm exp}\left[-\left(\frac{E}{E_c}\right)\right],
\end{equation}
where $N_{e,0}$ is the normalization constant of the electron-emitting spectrum, and $E_c$ is its cutoff energy.
The integrated electron spectrum is normalized to a total energy of $E_{CRe}=\int (dN_e/dE_{e})E_edE_e$ within the emission energies between  100\,MeV and about 43\,TeV, where the maximum energy is estimated from the equilibrium condition of acceleration gain and synchrotron loss for CR electrons \citep{2012ApJ...745..146K}. The photon spectrum of electron bremsstrahlung follows an analytical form given in \cite{1999ApJ...513..311B}, where the electron-nucleon and electron-electron processes are taken into account.

We use IC scattering as implemented in \texttt{Naima} package. The setup follows closely the framework demonstrated in \cite{2015ApJ...808...44F}. We used for the ISRF an averaged energy density of $0.57\,\rm eV\,cm^{-3}$\citep{2015ApJ...808...44F}, calculated from the LMC SED reported in \cite{2010A&A...519A..67I}. This includes the CMB ($\sim0.25\,\rm eV/cm^3$), IR ($\sim0.12\,\rm eV/cm^3$) and optical ($\sim0.32\,\rm eV/cm^3$) components for the radiation fields \citep{2022A&A...666A.167P}. The magnetic field used for the calculation of synchrotron spectrum is set to be $10\,\rm\mu G$, which is derived from the synchrotron intensity measurement at $4.8\,\rm GHz$ in the LMC \citep{2022MNRAS.510...11H}. We note this is a different value from what was used in \cite{2015ApJ...808...44F}, where they employ a mean magnetic field strength of $4.3\,\mu\rm G$, but this was derived for the magnetic field along the line of sight from Faraday rotation measures. Contribution from secondary CR electrons is omitted since they are only of $\sim10\%$ of the primary injected population \citep{2022A&A...666A.167P}. Lastly, the gamma-ray flux calculated from the pion decay component follows the prescription and parameterization given in \cite{2014PhRvD..90l3014K}. The number density of hydrogen atoms used here for the SGS\,4 region is fixed at a lower limit value of $\sim0.19\,\rm cm^{-3}$. It is derived from $n_{\rm H}=N_{\rm H}/l_{los}$, where $N_{\rm H}=0.7\times10^{21}\,\rm cm^{-2}$ is the hydrogen column density obtained from a X-ray analyses of SGS\,4 \citep{1994A&A...283L..21B,2001ApJS..136...99P}, and $l_{los}=1200\,\rm pc$ is the upper limit size of the SGS\,4 domain on the line of sight \citep{1980MNRAS.192..365M}. Similar to the case of CR electrons, we considered a PL and PLExp form for the emitting spectrum of parent protons, where the total energy of parent protons is normalized to $E_{CRp}$ for the energy range of 1\,GeV-1\,PeV. Here the maximum energy of 1\,PeV is sufficiently large to account for the highest-energy CR protons involved in the gamma-ray production analyzed here, and is consistent with the upper limit predicted by \cite{2001AstL...27..625B} for proton acceleration by an ensemble of shocks.

\begin{figure}[ht]
   \centering
   \includegraphics[width=1\hsize]{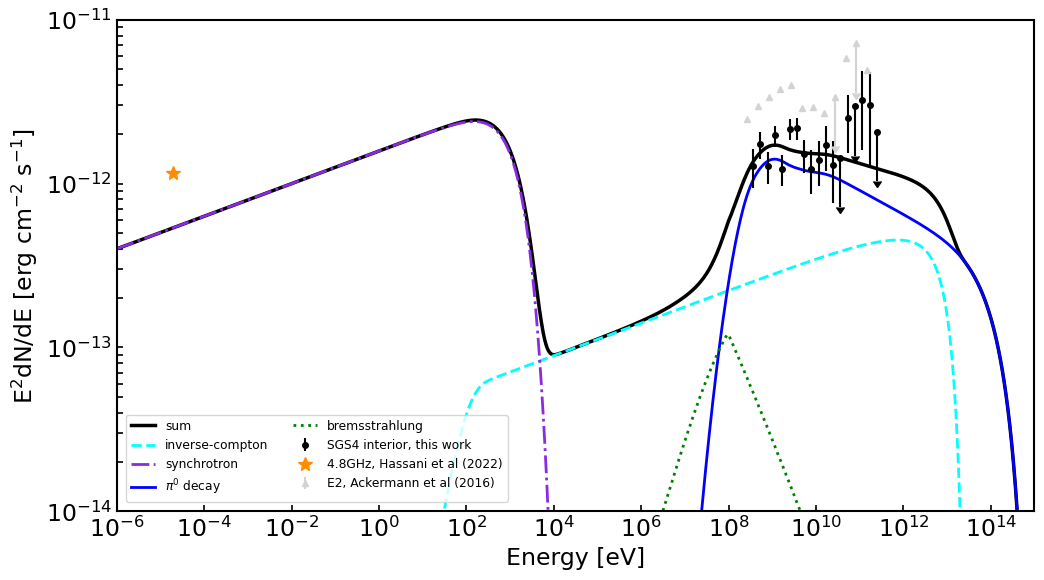}
      \caption{Broadband SED curves (solid black line) expected from the leptonic and hadronic models obtained with a radiation model for the interior of SGS\,4. The best-fitting SED curve is denoted with a black solid line. The associated components are plotted separately. The \textit{Fermi}-LAT data we analyzed are denoted with black markers, same dataset as presented in Fig.~\ref{fig:seds}. The E2 component (partly overlapped with SGS\,4) obtained in \cite{2016A&A...586A..71A} is plotted with light gray triangles for comparison. A rough estimate for the averaged radio synchrotron emission from SGS\,4 at 4.8\,GHz \citep{2022MNRAS.510...11H} is also shown here for comparison as an orange star.}\label{fig:sgs4_broad}
\end{figure}

We show in Fig.~\ref{fig:sgs4_broad} the resulting broadband spectrum for SGS\,4 interior region.
The best-fitting result corresponds to a parent particle population modeled with a PL spectrum, yielding a $\chi^2/n_{dof}=17.27/11$. There are four free parameters that enter the $\chi^2$ fit: the total energy in protons ($E_{CRp}$) and electrons ($E_{CRe}$), and their respective spectral indices, $\Gamma_p$ and $\Gamma_e$. For the electron population, we obtain an integrated energy of $E_{CRe}=(6.92\pm2.94)\times10^{51}\,\rm erg$ and a spectral index of $\Gamma_e=2.86\pm0.14$. A lower limit of $\Gamma_e\gtrsim2.6$ is imposed during the fitting, motivated by nonthermal synchrotron measurements reported in \cite{2022MNRAS.510...11H}. For protons, the best-fitting parameters are $E_{CRp}=(6.72\pm3.87)\times10^{52}\,\rm erg$ and $\Gamma_p=2.22\pm0.20$ (see also Table~\ref{tab:rad_model_paras} for an overview). 
\begin{table*}[ht]
\caption{Summary of the radiation model fits to the spectra of the SGS\,4 interior and the large-scale component in the LMC. }             % title of Table
\label{tab:rad_model_paras}    
\renewcommand{\arraystretch}{1.4}
\centering
\begin{tabular}{@{\extracolsep{5pt}}ccccccc}
%\begin{tabular}{l c c c c}
\hline\hline
 source &${\rm log_{10}}(E_{CRe}/\rm erg)$ & $\Gamma_e$ & ${\rm log_{10}}(E_c/\rm GeV)$ & ${\rm log_{10}}(E_{CRp}/\rm erg)$ & $\Gamma_p$ & ${\rm log_{10}}(E_c/\rm GeV)$\\
\hline
SGS4 interior & $51.84\pm0.18$& $2.86\pm0.14$& -& $52.98\pm0.18$& $2.22\pm0.20$&-\\
$\sqrt{\rm H\alpha}+160\,\rm\mu m$ &$52.64\pm0.13$ &$2.25\pm0.49$ &$2.62\pm0.44$ &$53.33\pm0.09$ &$2.40\pm0.12$ &$2.90\pm0.59$\\
\hline
\end{tabular}
\end{table*}
From the CR-induced individual emission component, we can observe that the gamma-ray radiation of SGS\,4 interior is largely dominated by the $\pi^0$ decay component at energies above few hundreds of MeV. Furthermore, in the absence of a cutoff in the CR spectra within the energy range considered for the parent particles, the resulting gamma-ray flux extends to several TeV, where the IC component becomes comparable and contributes the most to the total emission before declining rapidly as a result of the assumed maximum electron energy. After integrating each component within the energy range in our analysis, our result suggests the $\pi^0$ decay contributes a 76\%, IC a 23\%, and bremsstrahlung a 1\% to the overall energy fluxes in SGS\,4 interior. For energy flux measurements in radio (X-ray) waveband, we find one constraint (marked with an orange star) from the nonthermal synchrotron radiation\footnote{We make a rough estimate on the synchrotron emission of SGS\,4 region using Fig.~5 presented in \cite{2022MNRAS.510...11H}, where the estimated mean value of 25mJy/beam corresponds to $1.16\times10^{-12}\,\rm erg\,cm^{-2}\,s^{-1}$.} at a few GHz. As a comparison, we also plot the E2 component from \cite{2016A&A...586A..71A}, which partly overlaps with SGS\,4 interior and has an overall higher flux in comparison to SGS\,4 interior. This is because that the E2 component was characterized with a 2D Gaussian kernel of size $0.4^\circ$, covering a larger spatial extent than the spatial template of SGS\,4 interior used here, and thus would result higher integrated gamma-ray fluxes. The proton spectrum, with a power-law index of $\sim2.2$, is consistent with that resulting from diffusive shock acceleration.

\begin{figure}[ht]
   \centering
   \includegraphics[width=1\hsize]{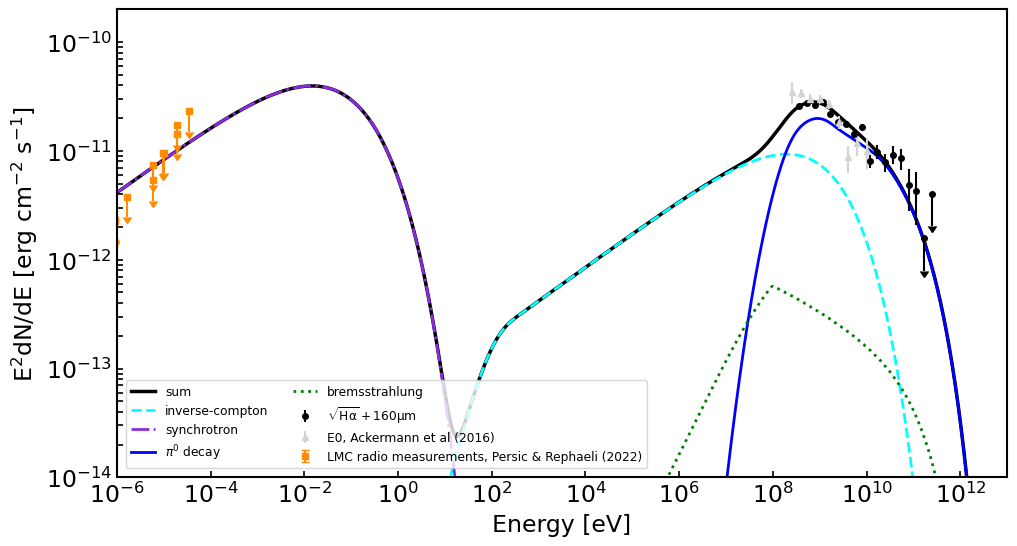}
      \caption{Same as Fig.~\ref{fig:sgs4_broad}, but for the combined large-scale components of $\sqrt{\rm H\alpha}$ and $160\mu\rm m$. The best-fitting SED curve is denoted with a black solid line. The light gray triangle error bars represent the large-scale disk component of E0 obtained in \cite{2016A&A...586A..71A}. Radio measurements on the LMC are shown as upper limits in orange (see \cite{2022A&A...666A.167P} for the collective radio data and corresponding references).}\label{fig:lmc_rad}
\end{figure}

On the other hand, we also study the spectrum of the total large-scale diffuse emissions derived from the combined $\sqrt{\rm H\alpha}$ and $160\mu\rm m$ templates (excluding the two largest star-forming regions of the LMC) with the same method employed in the case of SGS\,4. The broadband SED is plotted in Fig.~\ref{fig:lmc_rad}. 
The respective error bar in each bin is obtained via the standard uncertainty propagation of two SED datasets of $\sqrt{\rm H\alpha}$ and $160\mu\rm m$ templates. The best-fitting results correspond to a $\chi^2/n_{dof}=41.82/10$ with $n_{\rm H}=N_{\rm H}/l_{los}=1.13\,\rm cm^{-3}$ and PLExp for the model of parent particle spectrum. Cutoff energies of $\sim400\,\rm GeV$ and $\sim800\,\rm GeV$ are found for CR electrons and protons, respectively. The observed differential gamma-ray flux above a few hundreds of MeV can be largely attributed to a hadronic component. IC emission is comparable at energies around 100\,MeV. Bremsstrahlung component is negligible due to the low ion density employed here. It is important to note that our result in Fig.~\ref{fig:lmc_rad} shows only the large-scale diffuse emission as the contributions from each SGS region are accounted for with the corresponding SGS models. Integrating individual components within the selected LAT energy range, we obtain that the contribution from $\pi^0$ decay is about 76\%, IC a 22\%, and bremsstrahlung a 2\% to the overall energy fluxes across the LMC, excluding the 30 Dor and N11 star-forming regions. These values of contribution from each component change to 90\%, 8\%, and 2\% respectively when including 30 Dor and N11 star-forming regions in the large-scale templates of $\sqrt{\rm H\alpha}$ and $160\mu\rm m$. This differs from the percentages of results reported in \cite{2015ApJ...808...44F} or \cite{2017ApJ...843...42T} as the authors compute the overall energy fluxes emitted from the entire LMC instead of disentangling the large-scale diffuse regions from small-scale emission regions of SGSs. The integrated total photon flux over energy range of 200\,MeV-20\,GeV is calculated to be $\sim1.23\times10^{-7}\,\rm ph\,cm^{-2}\,s^{-1}$ using the best-fitting radiation models for all significantly detected SGSs and large-scale components in Table~\ref{tab:sedparas}, consistent with the integrated flux value ($\sim1.37\times10^{-7}\,\rm ph\,cm^{-2}\,s^{-1}$) reported in \cite{2015ApJ...808...44F}. The mild difference in the values result from contributions of point sources within the LMC boundary, which are accounted for and subtracted in our analysis. Our approach can more precisely label the mechanisms of CR-induced gamma-rays, and potentially distinguish the CRs living in SGSs from the ones in the large-scale disk of the LMC. This is also reflected in the injected proton indices in SGS\,4 and the large-scale diffuse component of the LMC, for example. The proton index of the large-scale region ($\Gamma_p=2.40\pm0.12$) is softer than that in the confined region of SGS\,4. This is expected due to the high CR injection rate and efficient confinement environment in the SGS regions \citep{2016A&A...586A..71A}.

\begin{figure}[ht]
   \centering
   \includegraphics[width=1\hsize]{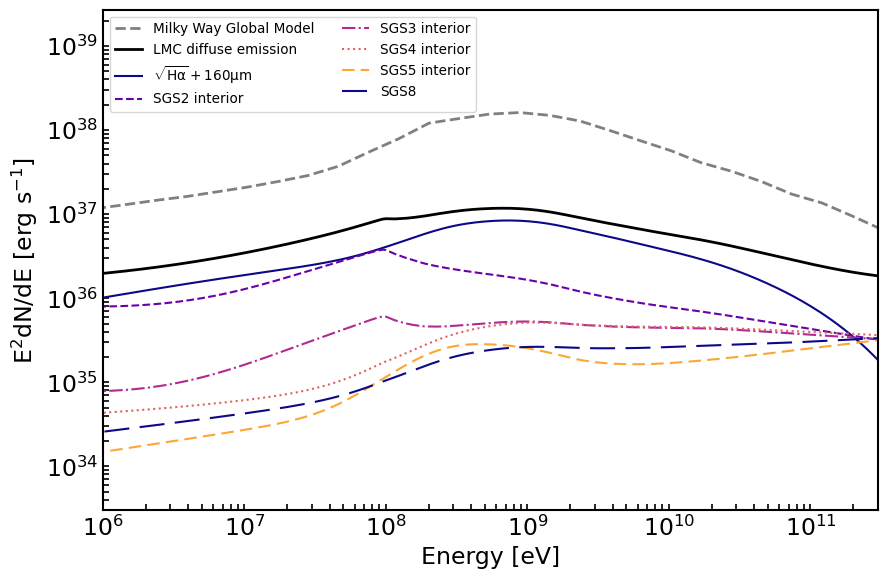}
      \caption{Gamma-ray luminosity spectra for different regions of the LMC. Black solid line denotes the sum of luminosities derived from best-fitting models of significantly detected sources listed in Table~\ref{tab:sedparas}. Spectrum of Milky Way (dashed gray line) calculated in \cite{2010ApJ...722L..58S} is plotted here for comparison.}\label{fig:lmc_all}
\end{figure}

The luminosities of CR-induced gamma-ray emission derived from the best-fitting radiation model for different regions of the LMC are plotted together in Fig.~\ref{fig:lmc_all}, where the black solid line represents the summed diffuse emission across the whole LMC. This is then compared with the Milky Way luminosity spectrum modeled in \cite{2010ApJ...722L..58S}.
The contribution from hard spectra of small-scale SGS regions leads to a mild hardening for the spectrum of the entire LMC diffuse emission at energies above $\sim$10\,GeV, as compared to the calculated Milky Way spectrum. The integrated luminosity over energy range of 1\,GeV-300\,GeV for the Milky Way spectrum plotted here is estimated to be $3.21\times10^{38}\,\rm erg/s$, which is about 10 times brighter than the integrated diffuse emission of the LMC for the same energy range.

The analyses exemplified in detail on SGS\,4 and the large-scale component show that the parent CR particles responsible for gamma-ray emissions have very different spectral indices and energy inputs. Also, despite the low ambient density of hydrogen atom in SGS\,4, the hadronic emission dominates over the selected energy range with a 76\% contribution to the total emission for its much-higher CR proton density, as compared to the sea of CRs in the LMC.  
%\section{Discussion}\label{dis}
\section{Discussion and conclusion}\label{sumandcon}
We presented a first study of the gamma-ray emissions toward the interior of SGSs in the LMC. With more than 15 years of \textit{Fermi}-LAT observations, we conducted a spatial analysis centered on LMC in the higher-energy band of 10\,GeV-300\,GeV to study the residual emissions, which we found to coincide noticeably with the locations of $\rm H\alpha$-identified SGSs. The spatial structure of the emission in the SGS interior was assumed to originate from the distribution of CR particles that are modulated by an ensemble of SNR shocks in SGSs. The fitting results presented in Table~\ref{tab:joint} indicate a non-Gaussian inner structure of emission for five significantly detected SGSs at higher energies. The parameters we used to establish the spatial templates of these SGSs varied from one case to the next. We will interpret them in a follow-up work. 

With the generated spatial templates for the interior of SGSs, we were able to extend the analysis to the full LAT range selected here, 300\,MeV-300\,GeV, and accounted for emissions from these small-scale SGS regions. For the diffuse emissions on the large scale in the LMC, we implemented various ISM gas and ISRF templates and assumed two extreme cases of the corresponding CR distributions: uniform, or confined to star-forming/particle-accelerator regions. We found that the combined templates of $\sqrt{\rm H\alpha}$ and $160\mu\rm m$ with uniformly distributed CR particles yielded the best significance for the analyzed ROI. Then, together with the isolated point-like sources detected and reported in the 4FGL-DR3 catalog, we separated the gamma-ray emissions from the three main components in the LMC: small-scale SGS regions, the large-scale region of the LMC, and the isolated sources. The small-scale SGS regions contribute an integrated energy flux of about $4.47\times10^{-11}\,\rm erg\,cm^{-2}\,s^{-1}$ to the entire LMC above 1\,GeV, which is $70\%$ of the energy emission from the large-scale components. However, it should be noted that the CR proton energy density calculated for the SGS regions is 3-10 times higher than that of the large-scale components ($\sim0.04\,\rm eV/cm^{-3}$), and it is 15\%-39\% of the local CR density for a value of $1\,\rm eV/cm^{-3}$ for the CR protons above 1\,GeV in the Milky Way. In addition, an upper limit on the cumulative emission from shell components of SGS\,2-5 and 7 was estimated to be $1.83\times10^{-11}\,\rm erg\,cm^{-2}\,s^{-1}$ at a confidence level of 95\% for energies above 1\,GeV.  

Particularly, we took SGS\,4 as an example to study the underlying particle distribution and emission/loss processes using the package \texttt{Naima}. The goodness of fit was acceptable, with a $\chi^2/n_{dof}=17.27/11$. The broadband SED model predicts for the analyzed energy range that a large fraction of 76\% of the energy emission is caused by the $\pi^0$ decay in the hadronic process, whereas IC only contributes 23\% to the total emission. In the other wavelengths, the projected synchrotron radiation curve and the radio observation at a few GHz agree reasonably well. Similarly, in the case of other SGSs with hard gamma-ray spectra (see Appendix~\ref{app:B} for, e.g., SGS\,2 and 3 in Figs.~\ref{fig:lmc2} and \ref{fig:lmc3}, respectively), the observed energy fluxes is largely attributed to the hadronic emission. In contrast, the gamma-ray spectra of SGS\,5 and 8 (shown in Figs.~\ref{fig:lmc5} and \ref{fig:lmc8}) are better interpreted by a dominant IC component, particularly at higher energies.  
The spectral index of the parent CR proton varies from 2.17 to 3.10 for the five SGSs that we fit with a broadband model. 
On the other hand, our derived broadband SEDs for some SGS regions can be compared to observations at higher energies. For example, the H.E.S.S. detection of R136 \citep{2024ApJ...970L..21A} reports a flux of $\sim3\times10^{-13}\,\rm erg\,cm^{-2}\,s^{-1}$ at 1\,TeV for a source size of $\sim0.04^\circ$. This can be compared to our measurement of the SGS\,2 interior ($\sim8\times10^{-13}\,\rm erg\,cm^{-2}\,s^{-1}$) for a size of $\sim0.5^\circ$. The other TeV source, N157B, which also spatially overlaps the SGS\,2 interior, is accounted for by 4FGL\,J0537.8-6909 from our source model in the analysis. 

These hard gamma-ray spectra of SGSs can be interpreted as originating from a CR population that is more highly energetic than the population that is distributed throughout the LMC. Alternatively, we note that it might be the result of a population of unresolved sources in the vicinity of SGSs. Measurements at other wavelengths are therefore necessary and crucial to further confirm the spectral features we observed here. In particular, future TeV-band observations of these SGS regions with, for example, the Cherenkov Telescope Array \citep{2023MNRAS.523.5353A} can help to elucidate the origin of the hard spectrum feature seen here.  

To study the spectral behavior of the large-scale component, we combined the two SED datasets from $\sqrt{\rm H\alpha}$ and $160\mu\rm m$ to obtain the total emission from the large-scale region of the LMC, and we then fitted it to the radiation model. The best-fitting result suggest a 76\% contribution to the total large-scale diffusion emission (excluding the 30 Dor and N11 regions) from the component of neutral pion-decay-induced emission, and a combined 24\% contribution from the leptonic process. The respective values changed to 90\% and 10\% when 30 Dor and N11 were included in the templates. The best-fitting model obtained here for the large-scale component can be further compared to the case when the spectrum of the LMC is extracted via various infrared survey maps (or combined), as was indicated in \cite{2020ApJ...894...88A}, who reported a correlation between the luminosities of gamma-ray and infrared measurements for star-forming galaxies.   
\begin{acknowledgements}
The \textit{Fermi} LAT Collaboration acknowledges generous ongoing support
from a number of agencies and institutes that have supported the
development and the operation of the LAT and a scientific data analysis.
These include the National Aeronautics and Space Administration and the
Department of Energy in the United States, the Commissariat \`a l'Energie Atomique
and the Centre National de la Recherche Scientifique / Institut National de Physique
Nucl\'eaire et de Physique des Particules in France, the Agenzia Spaziale Italiana
and the Istituto Nazionale di Fisica Nucleare in Italy, the Ministry of Education,
Culture, Sports, Science and Technology (MEXT), High Energy Accelerator Research
Organization (KEK) and Japan Aerospace Exploration Agency (JAXA) in Japan, and
the K.~A.~Wallenberg Foundation, the Swedish Research Council and the
Swedish National Space Board in Sweden.
Additional support for science analysis during the operations phase is gratefully 
acknowledged from the Istituto Nazionale di Astrofisica in Italy and the Centre 
National d'\'Etudes Spatiales in France. This work performed in part under DOE 
Contract DE-AC02-76SF00515.

\end{acknowledgements}

% WARNING
%-------------------------------------------------------------------
% Please note that we have included the references to the file aa.dem in
% order to compile it, but we ask you to:
%
% - use BibTeX with the regular commands:
%   \bibliographystyle{aa} % style aa.bst
%   \bibliography{Yourfile} % your references Yourfile.bib
%
% - join the .bib files when you upload your source files
%-------------------------------------------------------------------
\bibliographystyle{aa}
\bibliography{ref.bib}

\begin{appendix}
\section{Distribution of extended sources by \textit{Fermi}-LAT and H.E.S.S. in the LMC}\label{app:positions}
\begin{figure}[ht]
   \centering
   \includegraphics[width=0.8\hsize]{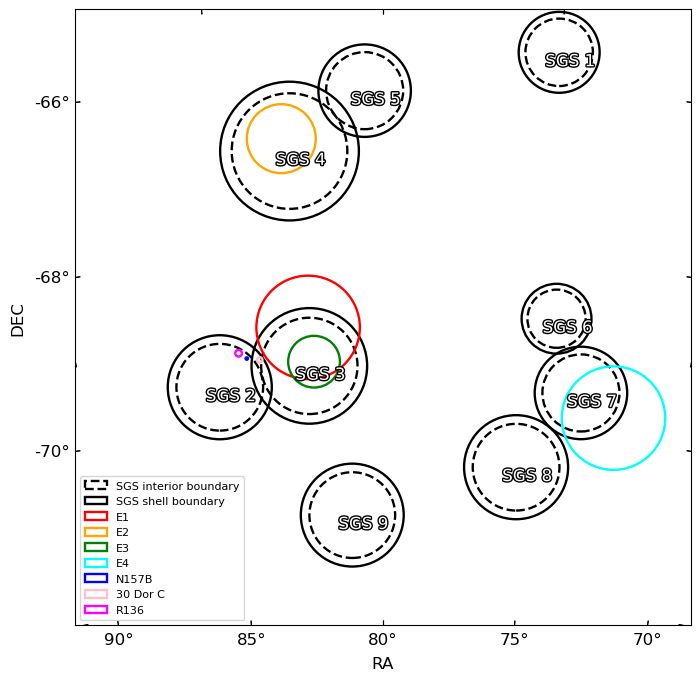}
   \caption{Locations of the H$\alpha$-selected SGSs \citep{1980MNRAS.192..365M} are marked with black circles (see Subsection~\ref{multiSNR} for more details on SGS shell and interior).  Extended sources (E1-E4) reported in previous \textit{Fermi}-LAT observations are also indicated \citep{2016A&A...586A..71A}. Relevant H.E.S.S sources (30 Dor C, N157B, R136) are marked as well \citep{2015Sci...347..406H,2024ApJ...970L..21A}.}\label{fig:positions}
\end{figure} 
\section{Residual TS map in the 10-300 GeV band}\label{app:resi_10gev}
\begin{figure}[ht]
   \centering
   \includegraphics[width=0.8\hsize]{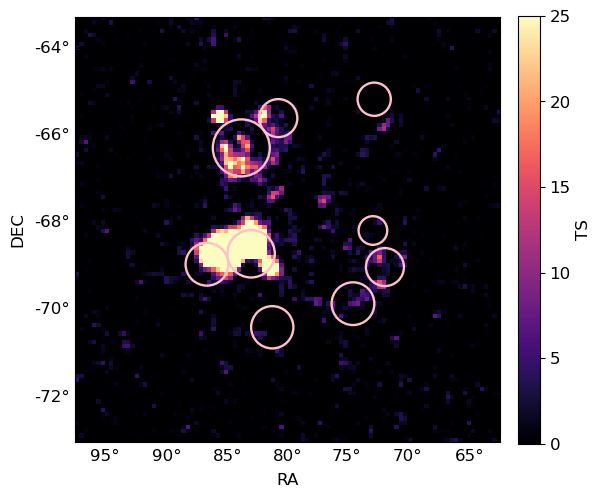}
   \caption{Residual TS map of the LMC in the 10-300\,GeV band.}\label{fig:resi_10gev}
\end{figure} 
\section{Other ISM, ISRF tracers} \label{app:others}
Under the hypothesis that large-scale gamma-ray emission across the LMC is due to uniformly distributed CR particles interacting with ISM gas and (or) ISRF, we list in Table~\ref{tab:otherism} the best log-likelihood fitting results when using different ISM/ISRF tracers for modeling gamma-ray emissions in the ROI.
\begin{table}[ht]
\caption{The best-fitting results for individual tracer template and combined templates with different CR distribution across the LMC.}             % title of Table
\label{tab:otherism}    
\renewcommand{\arraystretch}{1.4}
\centering
\begin{tabular}{ c|c|c } 
\hline
  & tracer & $l_{max,\,i}$ \\
\hline
 Background model & - & 68044.9 \\
\hline
\multirow{7}{11em}{Hypothesis 1: uniform CR} & HI & 80831.2 \\ 
& $\rm H_2$ & 81652.5 \\ 
& $\rm HI+\rm H_2$ (linear) & 80875.3 \\
& $\sqrt{\rm H\alpha}$ & 85883.6 \\ 
& $160\mu\rm m$ & 85786.8 \\ 
& $1.24\mu\rm m$ (starlight) & 78905.7 \\ 
& $\sqrt{\rm H\alpha}$, $160\mu\rm m$ & 86247.5 \\ 
\hline
\multirow{3}{11em}{Hypothesis 2: confined CR} & $\sqrt{\rm H\alpha}$ & 86099.6 \\ 
& $160\mu\rm m$ & 86026.6 \\ 
& $\sqrt{\rm H\alpha}$, $160\mu\rm m$ & 86122.5 \\ 
\hline
\end{tabular}
\tablefoot{References for templates: HI \citep{2016A&A...594A.116H}; $\rm H_2$ \citep{2016ApJ...825...12J}; H$\alpha$ \citep{2003ApJS..146..407F}; $160\mu\rm m$ \citep{2015PASJ...67...50D}; $1.24\mu\rm m$ \citep{2006AJ....131.1163S}. The spectral shapes for all templates are taken as a LogP form.}
\end{table}
In addition, we also list the best fitting results for an alternative hypothesis, under which the CR particles are confined locally near acceleration sites in the LMC. For this, we use locations of CR sources in the LMC as a tracer of the global CR spatial distribution. Diffusion of CR particles is taken into account using a Gaussian kernel of $0.2^\circ$ for simplification. 
Three populations of stellar objects are chosen for this purpose as candidate CR sources (or accelerators) in the LMC: the young massive stellar objects, X-ray SNRs, and WR stars (other data catalogs of objects, that is, pulsars, OB associations, are either incomplete or not available for us when preparing this draft).
\section{Residual significance map}\label{app:new_sig}
\begin{figure}[ht]
   \centering
   \includegraphics[width=0.8\hsize]{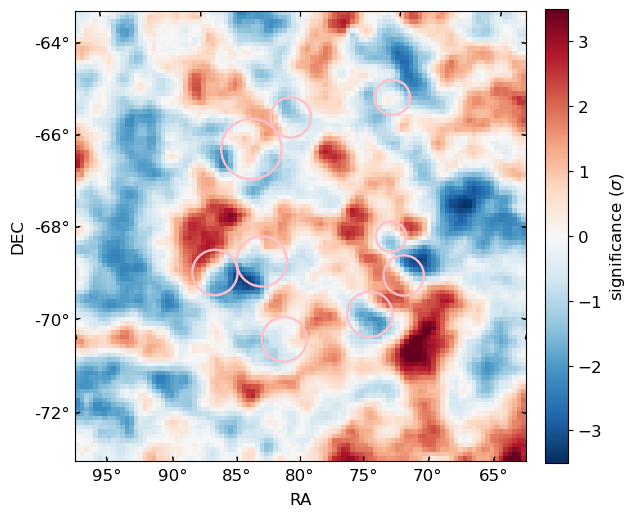}
   \caption{Residual significance map that corresponds to Fig.~\ref{fig:final_resimap}, showing the significance of the data-model residuals within the ROI. The color bar saturated at $\pm3.5\sigma$.}\label{fig:new_sig}
\end{figure} 
\section{Components of model 2}\label{app:compb_c}
\begin{figure}[ht]
   \centering
   \includegraphics[width=0.8\hsize]{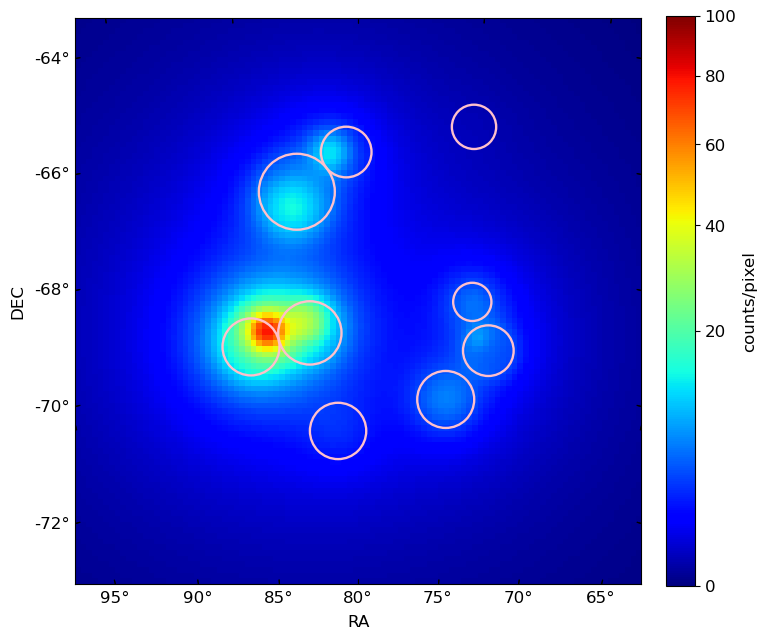}
   \includegraphics[width=0.8\hsize]{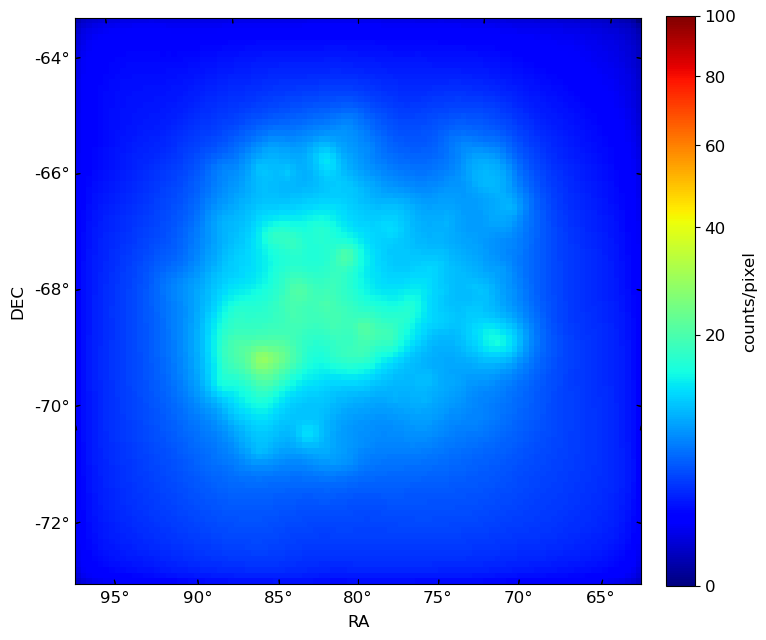}
   \caption{\textit{Top}: the corresponding Component B of model 2 for model map shown in Fig.~\ref{fig:final_resimap}; \textit{Bottom}: same as for left panel, but for Component C.}\label{fig:compb_c}
\end{figure} 
\section{Broadband SEDs of other SGSs}\label{app:B}
\begin{figure}[ht]
   \centering
   \includegraphics[width=1\hsize]{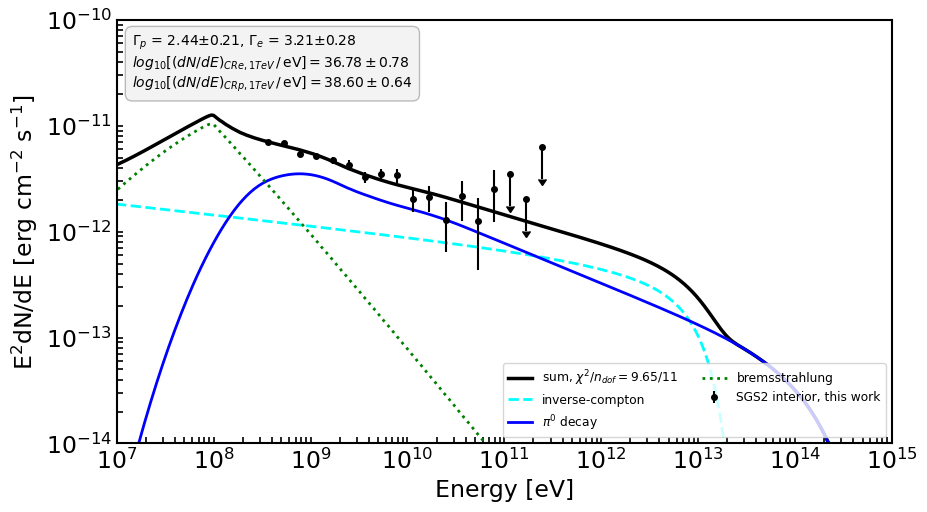}
      \caption{Same as Fig.~\ref{fig:sgs4_broad}, but for SGS\,2 interior. An ambient density $n_{\rm H}=1.33\,\rm cm^{-3}$ is used for SGS\,2 interior.}\label{fig:lmc2}
\end{figure} 
\begin{figure}[ht]
   \centering
   \includegraphics[width=1\hsize]{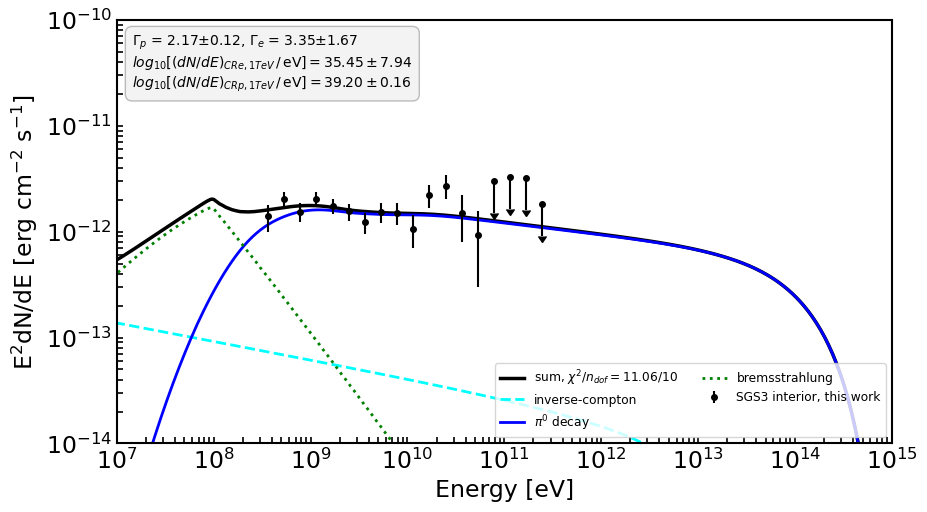}
      \caption{Same as Fig.~\ref{fig:sgs4_broad}, but for SGS\,3 interior. An ambient density $n_{\rm H}=0.49\,\rm cm^{-3}$ is used for SGS\,3 interior.}\label{fig:lmc3}
\end{figure} 
\begin{figure}[ht]
   \centering
   \includegraphics[width=1\hsize]{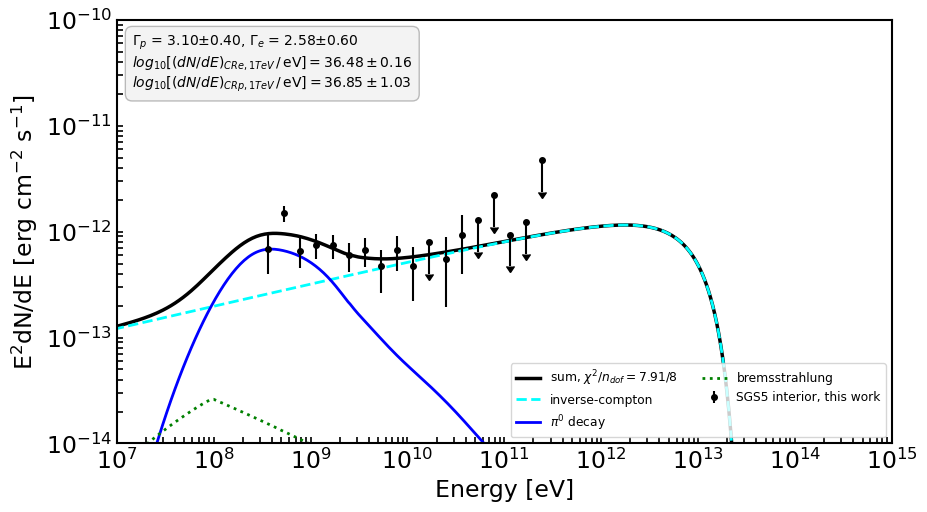}
      \caption{Same as Fig.~\ref{fig:sgs4_broad}, but for SGS5 interior. An ambient density $n_{\rm H}=0.48\,\rm cm^{-3}$ is used for SGS\,5 interior.}\label{fig:lmc5}
\end{figure} 
\begin{figure}[ht]
   \centering
   \includegraphics[width=1\hsize]{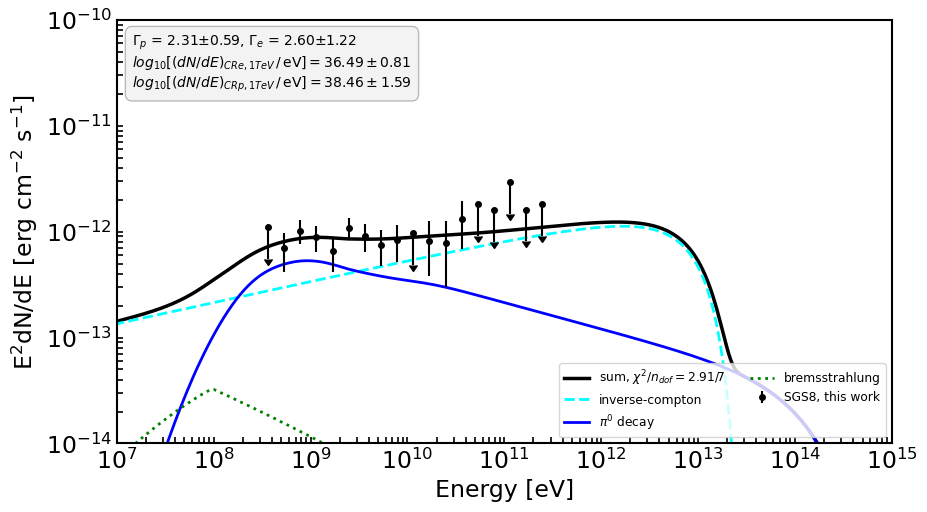}
      \caption{Same as Fig.~\ref{fig:sgs4_broad}, but for SGS\,8 (modeled with a Gaussian template). We assume an ambient density $n_{\rm H}=0.5\,\rm cm^{-3}$ for SGS\,8.}\label{fig:lmc8}
\end{figure} 

\end{appendix}
\end{document}